\documentclass[letterpaper,twocolumn,10pt]{article}
\usepackage{usenix-2020-09}

\usepackage{url}
\usepackage{booktabs}
\usepackage{adjustbox}
\usepackage{xspace}

\let\labelindent\relax
\usepackage{enumitem}
\usepackage{afterpage}

\usepackage{listings}
\usepackage{amsmath}
\usepackage{xspace}
\usepackage{tcolorbox}
\usepackage{listings}
\usepackage{xcolor}
\usepackage{titlesec}
\usepackage{enumitem}
\usepackage{algorithm}
\usepackage{algpseudocode}
\usepackage{adjustbox}
\usepackage{booktabs}
\usepackage{hyperref}
\usepackage[normalem]{ulem}
\usepackage{colortbl }
\usepackage{multirow}
\usepackage{hhline}
\usepackage{pdflscape}
\usepackage{pifont}
\usepackage{latexsym}
\usepackage{stmaryrd}
\usepackage{balance}
\usepackage{caption}
\usepackage{wasysym}
\usepackage[flushmargin,bottom]{footmisc}
\usepackage{comment}
\usepackage{tikz}
\usepackage{float}
\usepackage{amsmath}
\usepackage{amssymb}
\usepackage{dashrule}
\usepackage{verbatim}
\usepackage{authblk}
\usepackage[available]{usenixbadges}

\definecolor{dkgreen}{rgb}{0,.6,0}
\definecolor{dkblue}{rgb}{0,0,.6}
\definecolor{dkyellow}{cmyk}{0,0,.8,.3}

\newcommand{\code}[1]{\small{\texttt{\textcolor{black}{#1}}}\normalsize}

\newcommand{\textie}{\textit{i.e.,}\xspace}
\newcommand{\texteg}{\textit{e.g.,}\xspace}

\newcommand{\flow}{\textcolor{black}{\footnotesize$\rightarrow$\normalsize}}

\newcommand{\tablefix}[2]{\textcolor{teal}{+#1}\xspace\textcolor{red}{-#2}}

\newcommand{\issuelink}[1]{\href{https://github.com/php/php-src/issues/#1}{#1}}

\newcommand{\tablecode}[1]{\small\textcolor{black}{#1}\normalsize}

\newcommand*\circled[1]{\tikz[baseline=(char.base)]{\node[shape=circle,draw,inner sep=0.2pt] (char) {\textcolor{black}{#1}};}}

\newcommand{\name}{FlowFusion\xspace}
\newcommand{\totalbug}{158\xspace}
\newcommand{\pendbug}{18\xspace}
\newcommand{\expectbug}{4\xspace}
\newcommand{\confbug}{11\xspace}
\newcommand{\dupbug}{0\xspace}
\newcommand{\fixbug}{125\xspace}
\newcommand{\crashbug}{39\xspace}

\newcommand{\tests}{test cases\xspace}

\definecolor{darkred}{rgb}{0.55, 0.0, 0.0}

\begin{document}

%don't want date printed
\date{}

\title{\Large \bf Fuzzing the PHP Interpreter via Dataflow Fusion}

\author{{\fontsize{11.5pt}{11pt}\selectfont Yuancheng Jiang}}
\author{{\fontsize{11.5pt}{11pt}\selectfont Chuqi Zhang}}
\author{{\fontsize{11.5pt}{11pt}\selectfont Bonan Ruan}}
\author{{\fontsize{11.5pt}{11pt}\selectfont Jiahao Liu}}
\author{{\fontsize{11.5pt}{11pt}\selectfont Manuel Rigger}}
\author{{\fontsize{11.5pt}{11pt}\selectfont Roland H. C. Yap}}
\author{{\fontsize{11.5pt}{11pt}\selectfont Zhenkai Liang}}
\affil{\textit{{\fontsize{10.5pt}{11pt}\selectfont School of Computing, National University of Singapore}}}
\affil{\small\textsl{\{yuancheng, chuqiz, r-bonan, jiahao99, rigger, ryap, liangzk\}@comp.nus.edu.sg}\normalsize}

\maketitle

\thispagestyle{empty}

\begin{abstract}
%\boldmath
% \yc{official document call it ``PHP interpreter''}

% \yc{Page Limit: 13 Pages before Refs}

% \yc{PHP is in C (67\%), PHP (30\%), C++ only 0.6\%}

PHP, a dominant scripting language in web development, powers a vast range of websites, from personal blogs to major platforms. 
While existing research primarily focuses on PHP application-level security issues like code injection, memory errors within the PHP interpreter have been largely overlooked.
These memory errors, prevalent due to the PHP interpreter's extensive C codebase, pose significant risks to the confidentiality, integrity, and availability of PHP servers. 
This paper introduces \textit{\name}, the first automatic fuzzing framework to detect memory errors in the PHP interpreter. 
\name leverages dataflow as an efficient representation of test cases maintained by PHP developers, merging two or more test cases to produce fused test cases with more complex code semantics.
Moreover, \name employs strategies such as test mutation, interface fuzzing, and environment crossover to increase bug finding. 
In our evaluation, \name found \totalbug unknown bugs in the PHP interpreter, with \fixbug fixed and \confbug confirmed.
Comparing \name against the official test suite and a naive test concatenation approach, \name can detect new bugs that these methods miss, while also achieving greater code coverage. 
\name also outperformed state-of-the-art fuzzers AFL++ and Polyglot, covering 24\% more lines of code after 24 hours of fuzzing.
\name has gained wide recognition among PHP developers and is now integrated into the official PHP toolchain.

% The usefulness of \name has led to it being incorporated into the PHP repository.
% %
% We believe our approach is a practical and continuous way of enhancing the security of the PHP interpreter over time, given code changes along with their new test cases.

\end{abstract}

\section{Introduction}

\label{sec:intro}

PHP is a scripting programming language that is tailored for web development. Known for its flexibility and practicality, PHP powers a vast number of websites, ranging from personal blogs to global platforms. According to various reports~\cite{php1, php2, php3, php4}, PHP is used by over 70\% of websites worldwide, making it one of the most popular programming languages for web deployment.

% Over the past decade, PHP has maintained its dominance, consistently exceeding a 70\% usage share in terms of server-side programming languages for websites, as evidenced by \href{https://w3techs.com/technologies/history_overview/programming_language/ms/y}{usage statistics}. \cq{why not just cite [statistics] here?}

% \ry{other stats: https://benjamincrozat.com/php-is-dead-2024}

Although several PHP implementations are available, the official PHP interpreter is the most widely used. Its codebase, mainly in C, exceeds one million lines. As C is a low-level language without memory safety, the PHP interpreter is potentially vulnerable to memory errors that attackers may exploit.

% This work aims to enhance the security of the PHP ecosystem by reducing bugs that lead to memory errors.

% \footnote{These can be severe security issues. We quote from~\cite{chen2021one}: ``
% {\it
% Actually, the security team of Google also considers
% bugs in PHP interpreter as highly security-related.
% Therefore, our bugs in PHP interpreters, though not assigned with CVEs,
% can lead to severe security consequences
% }
% ''.}
% \ry{YC: see footnote}

% by malicious users,potentially leading to severe security impacts, including denial-of-service (DoS) attacks, mitigation bypasses, and even remote code execution. These risks highlight the critical need to remove memory error bugs from the for robust security measures or effective memory error detections to maintain the integrity and safety of the  PHP ecosystem. 
%
% However, because C is a memory-unsafe, low-level language, PHP is susceptible to memory errors.
% % such low-level code as shown by existing bugs can suffer from memory errors which affect security and robustness. 
% %
% Over the past three years, there have been over 500 publicly known memory errors. Such errors significantly degrade the robustness and security of applications built in PHP.

% \cq{Should we use ``we observed'' here, or say something like:
% "Within the past three years, XYZ publicly available runtime errors have been widely observed. Such errors degrade not only robustness but the stability of applications built upon PHP."}

Identifying memory errors in the PHP interpreter is challenging.
Existing research mainly focuses on script security issues, such as application bugs, including SQL injection and file inclusion.
%
% \cq{While the security of PHP has gained significant traction in recent times [cite], existing research mainly focuses on script security issues, such as application bugs, including SQL injection and file inclusion.}
%
Limited effort has been dedicated to detecting memory errors in the underlying PHP interpreter. 
To the best of our knowledge, no existing approach is specifically tailored for fuzzing the PHP interpreter. 
Some fuzzing approaches~\cite{holler2012fuzzing, aschermann2019nautilus, srivastava2021gramatron, chen2021one} rely on grammar-guided program generation and have found memory errors in the PHP interpreter. 
Nevertheless, they may not be sufficiently effective in finding such bugs because \textit{grammar generation focuses more on ensuring syntactic correctness than code semantics}. 
As a result, it struggles to create programs with complex semantic behavior and is unlikely to produce test cases targeting specific modules of the PHP interpreter. 

% Additionally, even with existing grammar efforts~\cite{}, 
% for example, grammar-guided generation may need to be updated with new PHP features requiring significant manual effort. Existing grammar-based approaches are unlikely to generate such specialized examples without special modeling. 
% \ry{changed}
% Nevertheless, PHP’s syntax is notably distinct from other programming languages, complicating the generation of diverse and valid inputs for fuzzing the PHP interpreter.  Therefore, these works require heavy manual effort and hardly touch deep and complex bugs requiring rich code semantics.

Previous wisdom in fuzzing research has proven its practicality in uncovering memory errors~\cite{han2018enhancing, ba2022efficient}. 
In terms of the PHP interpreter, the PHP community is maintaining a high-quality test suite consisting of over 19K test cases. Such test cases cover broad PHP features over 80 modules, such as in-memory databases or sessions, coming with plug-and-play running environments and configurations.

% The PHP interpreter has maintained a high-quality test suite consisting of over 19,000 official test cases.
% \cq{\^ I feel like there is a broken link. Why do you suddenly talk official test-cases? Start sth like:} 
% \cq{Previous wisdom in fuzzing research has proven its practicality in uncovering memory errors [cite some fuzzing paper in other memory-errors].
%
% Moreover, we observed that PHP has its unique advantage for fuzzing: its interpreter has maintained a high-quality test suite consisting of over 19K official test cases.}
% \footnote{Subsequently, if we refer to a test case (test), we refer to one test case of the official test suite unless otherwise noted.}
% These test cases are manually crafted, targeting specific PHP features like in-memory databases or sessions, coming with running environments and configurations.
% \cq{Such test cases cover broad PHP features over 80 modules, such as in-memory databases or sessions, coming with plug-and-play running environments and configurations}.
% This official test suite covers over 80 different modules, and running the test suite yields unexpectedly higher code coverage than all existing approaches.

\noindent\textit{Observation 1:} The test cases from PHP's official test suite yield higher code coverage than existing fuzzing approaches, thereby naturally forming a ``golden testbed'' for fuzzing.

Despite the intuition to leverage PHP official test cases for fuzzing, there still remains a key challenge---\textit{how to enrich code semantics of test cases to reveal memory errors?}
Specifically, even though the test cases are well-maintained with valid syntax, they are limited to simple code semantics given their unit-test-like nature.
Ideally, we could extend the official test suite to create a larger and more comprehensive test suite.
One approach is to craft semantically enriching code transformations, but doing so either remains inefficient if implemented as byte-level mutations or requires significant expertise and manual effort.
To develop a practical automated fuzzing strategy, we propose an alternative method of merging test cases. However, simply concatenating two existing tests is ineffective, as it yields the same outcome as running them independently.
To extend existing test cases into more effective ones, it is essential to capture and extend their code semantics. A standard approach is to extract the control flow and data flow of PHP programs in the official test suite.

% \zk{Readers are not ready to understand the fusion idea yet. Need to formulate the challenge, which is the requirement of complex semantics in test cases.}
%
% Furthermore\cq{Second,}, there is a lack of existing fuzzers for PHP, requiring us to handle the rich syntax of the PHP language from scratch, as no existing fuzzers are specifically tailored for PHP.
% \cq{\^\ This looks like an interpreterering issue to me instead of a research challenge. One solution, say something like: ``Given PHP language's complex properties of XYZ, it is unclear how to design an efficient fuzzer for it. Therefore, until now, there is no fuzzer...''} \bn{There are universal fuzzers, such as AFL, OSS-Fuzz and Honggfuzz. Do we need to explain why a specific fuzzer tailored for PHP is necessary here?}

\begin{figure}[t]
\begin{lstlisting}[language=PHP, label=lst:motivate, caption={A 20-year-old memory error found by \name}]
/* Test A */
(*@\textbf{\textcolor{teal}{/* Test A dataflow: \$dom $\rightarrow$ \$ref $\rightarrow$ \$nodes */}}@*)
  $dom = new DOMDocument;
  $dom->loadXML(..);
  $ref = $dom->documentElement->firstChild;
  (*@\underline{\textcolor{teal}{\$nodes}}@*) = $ref->childNodes;
  (*@\hdashrule{28em}{0.5pt}{2pt}@*)
/* Test B */
(*@\textbf{\textcolor{violet}{/* Test B dataflow: \$values $\rightarrow$ \$str $\rightarrow$ \$enc */}}@*)
  $values = array(..);
  foreach((*@\underline{\textcolor{violet}{\$values}}@*) as $str)
    { $enc = base64_encode($str); }
  (*@\hdashrule{28em}{0.5pt}{2pt}@*)
/* Our Fused Test */
(*@\textbf{/* Dataflow fusion:\textcolor{teal}{\$nodes} $\rightarrow$ \textcolor{black}{\$fusion} $\leftarrow$ \textcolor{violet}{\$values} */}@*)
  $dom = new DOMDocument;
  $dom->loadXML(..);
  $ref = $dom->documentElement->firstChild;
  (*@\underline{\textcolor{black}{\$fusion}}@*) = $ref->childNodes;
  $values = array(..);
  foreach((*@\underline{\textcolor{black}{\$fusion}}@*) as $str)
    { $enc = base64_encode($str); }
/* AddressSanitizer: heap-use-after-free */
\end{lstlisting}
\end{figure}

% To comprehensively capture the code semantics of the test case program, it is essential to consider both control flow and data flow together, as demonstrated in existing studies~\cite{liu2023learning}. 
% \cq{\^ broken link to me, why should you capture them? readers may not infer. Start sth like:}
% \cq{``To extend existing test cases into more effective ones, the essential way is to {\em capture and extend their code semantics}.
% %
% A standard approach towards that is to extract the program control- and data-flow.''}
%
% Nevertheless, our study reveals that 96.1\% of the tests exhibit sequential control flow, executing without branching. 

\noindent\textit{Observation 2}: Most official tests (96.1\%) exhibit sequential control flow (\textie no branches)---the code semantics of such programs can be effectively represented by mere dataflow.

%
% This finding suggests that control flow contributes little to the overall code semantics. Therefore, we recognize that the code semantics of the official test programs can be effectively represented using only dataflow. 

% We observe that 96.1\% of \tests exhibit sequential control flow (\textie \tests execute without branching) and recognize that these \tests can be well presented by \emph{dataflow}. \zk{the following breaks the flow of presenting ideas.} The dataflow is collected via dataflow analysis, \emph{understanding how data moves and transforms through different parts of a program by examining how variables are assigned, modified, and utilized}. 

% Although existing test cases are generally simple, some already generate complex data needed by other test cases, contributing to more intricate code semantics.\cq{I don't think this sentence is useful. What message does it convey? I would remove.}
%

% To address the challenge of generating complex code semantics, we use dataflow interleaving on the test cases to create new and interesting code interactions that were previously non-existent. 
To address the challenge of generating semantics-rich test cases, we use \textit{dataflow interleaving} to bridge the dataflow of two official tests, creating new code interactions (semantics) that were previously non-existent.
For example, in Listing~\ref{lst:motivate}, test A verifies DOM objects and their entity references, and test B checks base64 encoding. However, the official test suite does not account for more complex code semantics, such as encoding DOM-related objects using base64, which might occur in real-world scenarios. To this end, we combine these cases, alongside their semantics, thereby generating new code functionalities. We first present their dataflows consisting of dataflow A ([\code{\$dom}\flow \code{\$ref}\flow \code{\$nodes}]) and dataflow B ([\code{\$values}\flow \code{\$str}\flow \code{\$enc}]). Then, we interleave their dataflows by connecting a variable from dataflow A (\texteg \code{\$nodes}) to a variable from dataflow B (\texteg \code{\$values}). The connected variables are fused with a new one named \code{\$fusion}. As such, this real-world bug triggers an unknown 20-year-old use-after-free memory error (\textie it crashes PHP v5.0.0 released in 2004 and later versions) during the \code{foreach} iteration. By carefully crafting a malicious DOM, an adversary might exploit the memory vulnerability and compromise the host server supporting the PHP interpreter.
%
% Our goal is to enhance the official test suite by introducing test cases with more complex code semantics, which increases the test code coverage and uncovers hidden memory errors.\cq{Redundant to the first sentence. Be direct (and specific) to say what you do: ``To this end, we combine these cases, alongside their semantics, thereby generating new code functionalities.
% %
% This increases the test code coverage, but also uncovers hidden memory errors.''}

% By interleaving the dataflows of these tests, we can explore these more complex code semantics missed by the official test suite and enhance the testing coverage.

In this work, we present \textit{\name}, a novel approach to automate the discovery of memory errors in the PHP interpreter with the sanitizer oracle.\footnote{We use Address Sanitizer~\cite{serebryany2012addresssanitizer} (\textie ASan) and Undefined Behavior Sanitizer (\textie UBSan). Most of the ASan and UBSan reports belong to memory errors except arithmetic overflow.}
\name generates fused tests\footnote{Fused tests exhibit inherent diversity: combining pairs from the 19k official tests can produce over 300 million new tests.} which contain enriched code semantics from dataflow interleaving. We term our dataflow-driven test merging approach as \textit{dataflow fusion}. 
Additionally, \name employs several complementary strategies to more effectively uncover memory errors as follows.
% Apart from the aforementioned dataflow fusion, \name is built upon the following strategies
\begin{itemize}
[topsep=0.2mm,parsep=0.2mm,partopsep=0pt,leftmargin=*]
\setlength{\itemsep}{2pt}
\setlength{\parsep}{0pt}
\setlength{\parskip}{0pt}
    \item Test mutation.
    It mutates official test cases before their dataflows are interleaved.
    Test mutation is based on expression replacement, either replacing existing constants with special values, or replacing variables interchangeably.
    \item Interface fuzzing.
    It further makes use of more complex code semantics in fused test cases by calling random PHP functions with variables from fused test cases as arguments. 
    \item Environment crossover.
    It merges the execution environments (\texteg required modules, configurations) of fused test cases and further inserts random valid execution environment options collected from the official test suite.
\end{itemize}

We conducted experiments to assess the effectiveness of our approach. Remarkably, we detected \totalbug unique and previously unknown bugs in the PHP interpreter, of which \fixbug have been fixed and \confbug confirmed. These bugs span 10 different \emph{Common Weakness Enumerations} (CWEs) and affected over 80 individual source files in the PHP interpreter with the fixes changing over 5k lines of code.

We compared the effectiveness of \name against the official \tests and a simple concatenation approach of these \tests. The results demonstrate that \name not only detects more memory errors but also achieves higher code coverage. Additionally, we compared \name with state-of-the-art fuzzing techniques known for uncovering memory errors in the PHP interpreter. Specifically, \name outperformed AFL++\cite{fioraldi2020afl++} and PolyGlot\cite{chen2021one} by covering 24\% more lines of code after 24 hours of fuzzing under the same execution conditions.

We performed an ablation study of our primary approach—dataflow fusion—alongside other strategies such as test mutation, interface fuzzing, and environment crossover. The results provide a deeper understanding of how each strategy contributes to our approach's overall effectiveness. \name is recognized by the PHP developers and has been integrated into the official toolchain.\footnote{\name is available at \url{https://github.com/php/flowfusion}}
% We believe dataflow fusion can be applied to other programming language implementations to effectively fuse test cases.

In summary, we make the following contributions:

\begin{itemize}
[topsep=0.2mm,parsep=0.2mm,partopsep=0pt,leftmargin=*]
\setlength{\itemsep}{2pt}
\setlength{\parsep}{0pt}
\setlength{\parskip}{0pt}
    \item We propose a novel approach to discover memory errors in the PHP interpreter, called dataflow fusion. Based on the good quality and quantity of official test cases in the PHP interpreter, we expand the test suite with fused test cases in new code semantics by interleaving their dataflows.
    \item We implement our approach along with test mutation, interface fuzzing, and environment crossover as the first automatic fuzzing framework, \name, which effectively achieves comprehensive coverage in the PHP interpreter.
    \item \name has proven effective in discovering new bugs in the PHP interpreter. In total, we identified \totalbug unknown bugs, of which \fixbug have been fixed and \confbug confirmed.
\end{itemize}

\section{Background}

\label{sec:background}

% \begin{figure}[b]
%     \centering
%     \includegraphics[scale=0.33]{figures/phparch.pdf}
%     \caption{Architecture of the PHP interpreter}
%     \label{fig:phparch}
% \end{figure}
In this paper, the PHP interpreter refers to the official implementation.\footnote{\url{https://github.com/php/php-src}} 
It comprises three primary components: the Zend engine (\textie the \textit{zend} directory) is PHP’s core execution engine, encompassing the bytecode compiler, runtime executor, and memory management routines. It transforms PHP scripts into opcodes and executes them. The core modules (\textie the \textit{ext} directory) contain PHP’s built-in and bundled extensions (\texteg session, sqlite3), with each subfolder providing source code that extends the language beyond the core engine. The main functions (\textie the \textit{main} directory) hold essential PHP runtime and infrastructure code that initializes and orchestrates the interpreter, including entry points, configuration loading, and the main execution loop.

\vspace{+2mm}
\textbf{Memory errors in the PHP interpreter}.
\label{sec:memory_error}
% Statistics show that PHP ranks top 50 in 50 vendors with the most vulnerabilities~\cite{cvedetailsVendorsTotal}.
Statistics show that PHP ranks top 50 in the ranking for projects with the most vulnerabilities~\cite{cvedetailsVendorsTotal}.
%
% Since the PHP interpreter is written in C
Given its complex codebase and written in the memory-unsafe language (\textie C), it is prone to memory errors such as buffer overflows and use-after-free. We conducted a study of public issues reported from 2022 to 2024 in the official PHP GitHub repository, up to August 2024. Out of 567 verified and closed issues, we identified 191 as memory errors, representing a significant proportion (33.7\%) of all resolved bugs. 

To better understand the root cause of these memory errors, we conducted a triage analysis for the 191 reported bugs. Due to the lack of issue normalization and diverse bug descriptions, we manually categorized these bugs into 10 categories based on the issue reports, conversations, and patches. 
%
% The result indicates that \textit{Null Dereference}, \textit{Memory Leak or Exhaustion}, and \textit{Buffer Overflow or Underflow} are the three most common memory errors in PHP interpreter, accounting for more than 10\% each, except for other uncategorized miscellaneous bugs. 
The result indicates that \textit{null dereference}, \textit{memory leak or exhaustion}, and \textit{buffer overflow or underflow} are the three most common memory errors in PHP interpreter, accounting for more than 10\% each. 
In addition, \textit{use-after-free} bugs also occur occasionally, reaching 7\%. These bugs can lead to the aforesaid various security risks. In particular, \textit{use-after-free} and \textit{buffer overflow} ranked first and second respectively in the ``2023 CWE Top 10 Known Exploited Vulnerabilities (KEV) Weaknesses'' list~\cite{mitre2023Weaknesses}, indicating their high exploitability and the urgency of detection.

% \paragraph{Memory Error Detection for the PHP Interpreter}

% \label{sec:2.3}

% Limited research has been devoted to identifying memory errors in the PHP interpreter. Instead, the security community has focused extensively on detecting application vulnerabilities (\texteg SQL injection) within PHP scripts. These efforts are entirely different from our approach, as we aim to enhance the underlying robustness of the PHP interpreter. 
% LangFuzz~\cite{holler2012fuzzing}, NAUTILUS~\cite{aschermann2019nautilus}, Gramatron~\cite{srivastava2021gramatron}, and PolyGlot~\cite{chen2021one} are existing fuzzing approaches that have found memory errors in the PHP interpreter. These approaches have found memory errors in PHP by generating PHP programs and rely on significant manual effort in creating PHP grammar to ensure syntax correctness. Additionally, these approaches focus on the scalability of grammar usage and none is designed specifically for fuzzing the PHP interpreter, thus, showing limited effectiveness in finding memory errors in the PHP interpreter. 

% Other lesser-known but specific efforts, \yc{shall we mention non top tier papers?} such as PHPIL\cite{rao2020phpil} and PHP Fuzz\cite{baumgarte2023fuzzing}, use bytecode generation or sample code generation to fuzz the PHP interpreter. However, these approaches have a small impact in uncovering new memory errors in the PHP interpreter. 

\begin{figure}[t]
\begin{lstlisting}[escapeinside={(*}{*)}, language=PHP, label=lst:phpt, caption=Example test case in the official test suite, aboveskip=-2pt, belowskip=-3pt]
--TEST--
  FFI 007: Pointer comparison
--EXTENSIONS--
  ffi
--INI--
  ffi.enable=1
--FILE--
  <?php
    $ffi = FFI::cdef();
    $v = $ffi->new("int*[3]");
    $v[0] = $ffi->new("int[1]", false);
    $v[1] = $ffi->new("int[1]", false);
    $v[2] = $v[1];
    $v[1][0] = 42;
    var_dump($v[0] == $v[1]);
    var_dump($v[1] == $v[2]);
    FFI::free($v[0]);
    FFI::free($v[1]);
  ?>
--EXPECT--
  bool(false)
  bool(true)
\end{lstlisting}
\end{figure}

\vspace{+2mm}
\textbf{The official PHP test suite}.
\label{sec:test_suite}
The PHP community maintains the official test suite, which aids in verifying both the correctness and robustness of the PHP interpreter. The PHP test suite has over 19k distinct test cases, encompassing a wide range of code semantics with valid syntax. These test cases cover over 80 unique modules in the PHP interpreter and integrate with all existing bug-triggering reproducers as additional security test cases. While the overall test suite has diverse code semantics, the individual test case only verifies the functionality or the correctness of a single component. 
Therefore, 96.1\% of these test cases exhibit sequential control flow and execution without branches. Listing~\ref{lst:phpt} shows one example test case, which verifies the correctness of pointer comparison in the Foreign Function Interface (FFI) component of the PHP interpreter. 

The official test cases are in a special format consisting of sections delimited by \code{-{}-section-{}-} and formatted in \code{.phpt} file. It contains over 30 sections\footnote{\url{https://qa.php.net/phpt_details.php}} representing different meanings. Referring to Listing~\ref{lst:phpt}, we introduce some important sections as follows: (i) \code{-{}-test-{}-} section is a brief description of the test; (ii) \code{-{}-extensions-{}-} section details the extensions required for the test; (iii) \code{-{}-ini-{}-} section gives the specific configurations needed; (iv) \code{-{}-file-{}-} section contains the PHP program; and (v) \code{-{}-expect-{}-} section specifies the expected test results.

% \ry{add from rebuttal Question}
% The various test cases are in the PHP repository organized in the \verb+tests+ subdirectory within a module, e.g. Zend tests are in \verb+php-src/tree/master/Zend/tests+.

\vspace{+2mm}
\textbf{Dataflow of PHP programs.} 
\label{sec:dataflow}
Dataflow analysis~\cite{khedker2017data} involves tracking how data values propagate and are manipulated across different parts of a program. This includes identifying which variables are defined, used, or modified at different points in the code and understanding how these changes impact program behavior. 

Given a node $n$ in the control flow graph of the program, dataflow analysis considers the following four properties: the Gen Set (GEN$(n)$) is the set of definitions generated (\textie initialized) at a particular node $n$; the Kill Set (KILL$(n)$) is the set of definitions that are killed (\textie overwritten or invalidated) by the execution of a node $n$; the In Set (IN$(n)$) represents the set of dataflow facts coming into of a node $n$; the Out Set (OUT$(n)$) represents the set of dataflow facts going out of a node $n$.
% We next explain how we infer dataflow in the PHP programs using these concepts.

% \begin{equation}
% \label{eq:1}
%     \text{IN}(n) = \bigcup_{p \in \text{pred}(n)} \text{OUT}(p)
% \end{equation}
% \begin{equation}
% \label{eq:2}
%     \text{OUT}(n) = \text{GEN}(n) \cup \left( \text{IN}(n) \setminus \text{KILL}(n) \right)
% \end{equation}
% \vspace{-3mm}

% Equation~\ref{eq:1} and Equation~\ref{eq:2} are two basic principles widely used in dataflow analysis, which is first to calculate outgoing states of the first node $n_0$, then start state propagation (\textie IN($n_0$)\flow OUT($n_0)$\flow IN($n_1$)\flow OUT($n_1$)\flow..) in the following nodes. 

% In these equations, the $n$ usually represents a node in the control flow graph (CFG). The Gen Set (GEN$(n)$) is the set of definitions generated (\textie initialized) at a particular node $n$; the Kill Set (KILL$(n)$) is the set of definitions that are killed (\textie overwritten or invalidated) by the execution of a node $n$; the In Set (IN$(n)$) represents the reaching state\footnote{A reaching state is a node of a program where a particular variable definition remains active and can potentially influence subsequent computations.} of variables and expressions at the entry of a node $n$; the Out Set (OUT$(n)$) represents the state of variables and expressions available at the exit of a node $n$. The pred$(n)$ refers to the set of predecessor nodes of $n$ in the control flow graph.

In this work, we treat each statement in the test program as a node and denoted as $n_i$, where $i$ indicates the sequence of statements. Given a program P with N statements, we consider the P's dataflow as the collection of In Sets and Out Sets from all N statements, as shown in Equation~\ref{eq:1}. Based on these sets of dataflow facts, one can infer the dataflow graph (we omit it for simplicity) of the program. 

\vspace{-1mm}
\begin{equation}
    \label{eq:1}
    \text{Flow}(P) = \{ \text{IN}(n_i), \text{OUT}(n_i) \mid \forall i \in N \}
\end{equation}

For calculating the In Set, considering the sequential control flow of test programs, only the preceding statement has the coming-in dataflow facts. In Equation~\ref{eq:2}, the In Set containing the dataflow facts coming in is represented as the empty set when the first statement, and propagated from going out dataflow facts of preceding statement otherwise.

\vspace{-1mm}
\begin{equation}
\label{eq:2}
    \text{IN}(n_i) = \emptyset \; \text{ if } \; i=0 \; \text{ else } \; \text{OUT}(n_{i-1})
\end{equation}

To calculate the Out Set, we consider the definitions that are newly generated or eliminated, as represented by the Gen Set and the Kill Set. For dataflow facts related to function calls, the conventional approach is to either perform dataflow analysis within the function or utilize function summaries. Our method assumes that return values have data dependencies on function arguments for better efficiency. We notate these dataflow facts using the Fun Set (FUN$(n)$). Accordingly, we express the Out Set as shown in Equation~\ref{eq:3}. Specifically, the outgoing dataflow facts comprise elements from the incoming set, the generated set, and the function call set, while excluding those definitions that have been killed.

\vspace{-2mm}
\begin{equation}
\label{eq:3}
    \text{OUT}(n_i) = \text{GEN}(n_i) \cup \left( \text{IN}(n_i) \setminus \text{KILL}(n_i) \right) \cup \text{FUN}(n_i)
\end{equation}

In this paper, we use the notation \code{([$v_1$]\flow [$v_2$]\flow $\ldots$ [$v_{m-1}$]\flow [$v_m$])} to concisely represent dataflow from variable $v_1$ to $v_m$ in PHP programs.

\section{Threat Model}

\label{sec:threat_model}

Memory errors pose inherent security risks and frequently lead to vulnerabilities in the PHP interpreter. According to vulnerability statistics for PHP,\cite{php-cve-info}, more than 80\% (179/220) of PHP CVEs in the past decade resulted from overflows or memory corruptions. Google OSS-Fuzz~\cite{serebryany2017oss} project also considers PHP an interesting target for finding memory errors. Malicious users can attack PHP servers in the following ways:

\textit{(a) Malicious input interactions.} Attackers may craft specific inputs to PHP applications, resulting in direct user data leakage. For instance, there are interfaces from officially disclosed vulnerabilities, one heap buffer over-read~\cite{php-heap-oob} in the \textit{mysqlnd} PHP extension and another overflow~\cite{php-phar-oob} in \code{phar\_dir\_read()}. With such vulnerable interfaces, the attackers are allowed to craft SQL queries or \textit{phar} files to retrieve data. 
% can leak users' information via normal interactions in web applications like SQL queries or \textit{phar} files, 
% further highlighting these risks. 
% Such vulnerabilities can be controlled by user input and the exploitation route is typical to web applications.
Such memory errors are linked to user inputs and can be exploited in common web applications via malicious data.

\textit{(b) Malicious code injections.} Another way for attackers is to exploit non-interactive memory errors by using custom PHP script execution. Such code injections are common in online PHP editors or coding sandboxes. In these environments, users can run arbitrary code yet are prevented from executing system commands, since dangerous functions are usually disabled on well-maintained servers. Although under such restricted conditions, memory errors offer an alternative attack surface to bypass security measures or mitigation~\cite{tarlogicDeepDive, hackeroneInternetBounty}. Previous work, Polyglot~\cite{chen2021one}, provides a detailed example of how a memory error can be exploited to escape PHP sandboxes.

\section{Approach}

\label{sec:approach}

\begin{figure*}[t]
\setlength{\abovecaptionskip}{5pt}
\setlength{\belowcaptionskip}{0pt}
\setlength{\intextsep}{0pt}
    \centering
    \includegraphics[scale=0.76]{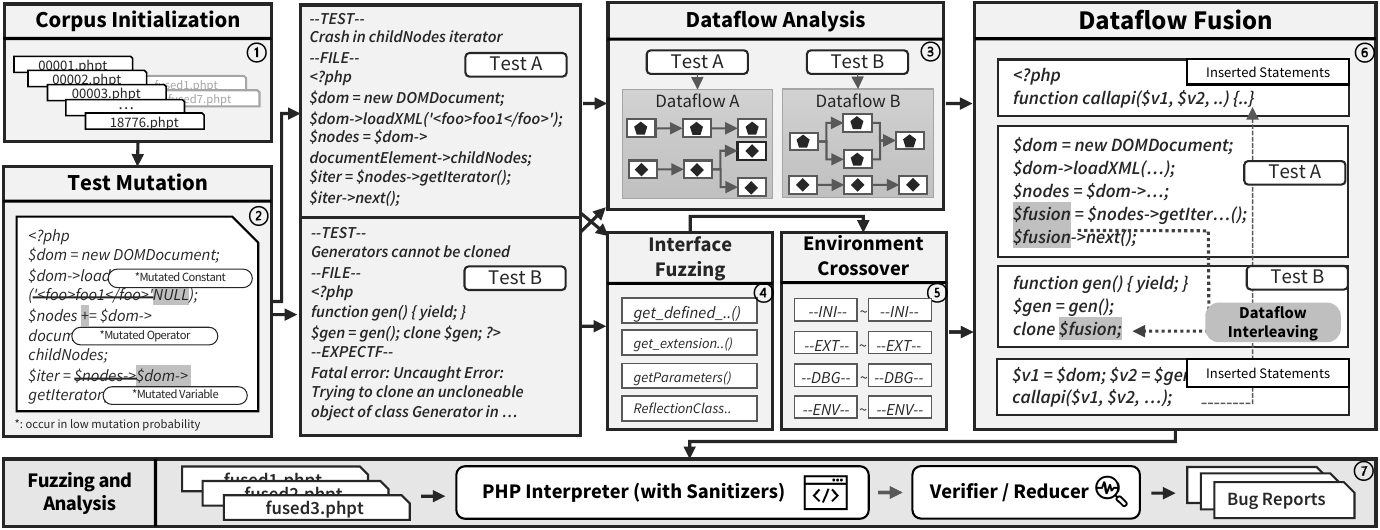}
    \caption{Approach Overview --- Illustrated with Segfault Error Found by \name in the Zend Engine}
    \label{fig:overview}
\end{figure*}

In this section, we first present an overview of our approach in Section~\ref{sec:approach_overview}, outlining the multiple steps to detect memory errors in the PHP interpreter. 
Then, we provide a detailed explanation of our key method, dataflow fusion, in Section~\ref{sec:algorithm}. Additionally, we describe other important strategies employed in this work to facilitate memory error detection in Section~\ref{sec:strategies}.

\vspace{-2mm}
\subsection{Approach Overview}

\label{sec:approach_overview}

Our approach is designed to uncover memory errors in the PHP interpreter. We utilize existing official test cases as seed programs to generate numerous new fused tests for continuous automatic fuzzing. The advantages of reusing the official test suite are twofold: (i) these tests are well-maintained over time with valid grammar and rich semantics, and (ii) their large quantity provides a substantial basis for test generation. By combining two or more tests, one can create a very much larger set of test cases which can also inherit the quality of the original tests. Our approach is inherently compatible and sustainable with the continuous addition of official test cases by PHP developers, which can automatically cover new features and bugs, ensuring more comprehensive testing.

We designed our approach with the following seven steps: (1) corpus initialization, (2) test mutation, (3) dataflow analysis, (4) interface fuzzing, (5) environment crossover, (6) dataflow fusion, and (7) result analysis. We present the overview of our approach in Figure~\ref{fig:overview} with a real unknown segmentation fault\footnote{\url{https://github.com/php/php-src/issues/14741}} found by \name.

% More details are in: Section~\ref{sec:3.3} and~\ref{sec:3.4}. The overview in Figure~\ref{fig:overview} illustrates 

% \yc{if change constructs, figure needs update}
At step~\circled{1}, \name initializes the test corpus with the 19k unique test cases from the official test suite. We first randomly select two seed tests from the corpus (for simplicity, we consider two tests, but more can be used).
At step~\circled{2}, \name applies test mutations to these chosen tests. The key idea behind test mutation is to introduce additional randomness or perturbations (in a low probability) to the seed tests before they are fused. This is done by interchangeably replacing operands or expressions, or substituting them with special values. 

At step~\circled{3}, \name employs a variable-level dataflow analysis on the program of each test. In Figure~\ref{fig:overview}, \name finds the dataflow ([\code{\$dom}\flow \code{\$nodes}\flow \code{\$iter}]) in test A and the dataflow ([\code{gen()}\flow\code{\$gen}]) in test B. The extracted dataflows will be used at step \circled{6} later. Next, \name prepares statements for fuzzing interfaces provided by the PHP interpreter at step~\circled{4}. It reuses variables from the fused code semantic as the arguments to call random functions implemented by PHP developers. We insert statements above and below the fused test to define random function calls, randomly select variables as arguments, and invoke function interfaces accordingly. 

At step~\circled{5}, \name first merges the execution environments of seed tests and then inserts random configurations from the test suite. As described in Section~\ref{sec:test_suite}, the official test cases include not only the PHP program but also other sections, such as required modules and configuration settings. 
Environment crossover begins by merging other sections of two seed test cases and then randomly inserts random configurations that are collected from the official test suite. This process ensures that the prerequisite conditions for fused tests are met by combining these additional sections, while also introducing variability into the execution environments through random configuration insertions. 

At step~\circled{6}, dataflow fusion links some dataflows from test A with test B to connect two seed tests. The key insight is to interleave dataflow by connecting variables across seed tests. 
As shown in Figure~\ref{fig:overview}, our approach creates a new test by replacing variable \code{\$iter} and \code{\$gen} with a bridging variable \code{\$fusion} by the guidance of extracted dataflow in the previous step. The code semantic is changed accordingly, now cloning the DOM node rather than previously a function object, which is not tested in the official test suite. Details of dataflow fusion are explained in Section~\ref{sec:algorithm}.

Finally, in step~\circled{7}, after our approach completes dataflow fusion, the PHP interpreter is invoked to execute the fused test cases and check for any sanitizer violations. We rely on the sanitizer oracle (\textie ASan and UBSan) to detect bugs. Note that although we focus on detecting memory error bugs, as we are using an undefined behavior sanitizer, we also catch other bugs like integer overflows. 
The execution generates logs of standard inputs and errors, as well as reproducing scripts. We use these outputs to deduplicate crashes by checking the crash site and mapping the backward stack trace. Once verified, \name reduces the reproducer of sanitizer violations to a minimal version using delta debugging~\cite{zeller2002simplifying}. Last, we assess the potential security impact of the discovered bugs and follow the PHP community's disclosure policy:
% \ry{YC: if need space can move to footnote}
(i) reporting low-severity bugs (\texteg null pointer dereference) directly at issues and (ii) reporting high-severity bugs (\texteg stack overflow, heap use-after-free) at the security page, where vulnerability reports remain private until analyzed and fixed. 

\subsection{Dataflow Fusion}

% \cq{I'm a bit confused about the current section layout (3.2 - 3.4): In 3.2's overview, you mentioned Steps 5 (environment crossover) and 6 (dataflow fusion) are key steps. However, you introduce data flow fusion in 3.3, but put environment crossover into a heading of 3.4 ``Further strategies''. This looks less systematic.}
% \ry{this is mainly for space}

% \cq{One idea, say step 4,5,6 are three challenges. Then use 3 sub-sections to tackle them?}
% \ry{not really challenges, this is an approach, the challenge is generating effetive tests to uncover memory errors}

\label{sec:algorithm}

We introduce our core insight, dataflow fusion, by referring to Algorithm~\ref{alg:fusion}. Dataflow fusion aims to combine two programs, $A$ and $B$, extracted from two random seed tests, into a single fused program $F$. The process begins by calculating the IN and OUT sets of dataflow facts for each program using the \texttt{ComputeDataflowSets} function. This function initializes the IN and OUT sets for each node within a program based on GEN and FUN sets and iteratively computes these sets using Equation~\ref{eq:2} and Equation~\ref{eq:3}. Specifically, for each node $n_i$, the IN set is derived from the OUT set of the preceding node, while the OUT set is computed using a union of GEN, IN, and FUN sets minus the KILL set. 

\begin{algorithm}
\caption{Pseudocode of Dataflow Fusion}
\begin{footnotesize}
\label{alg:fusion}
\begin{algorithmic}[1]
\State \textbf{Input:} programs $A$, $B$ extracted from two random seed tests 
\State \textbf{Output:} fused program $F$

\Function{ComputeDataflowSets}{$P$}
    \State Initialize $\text{IN}(n_1) = \emptyset$ and $\text{OUT}(n_1) = \text{GEN}(n_1) \cup \text{FUN}(n_1)$
    \For{$i = 2$ to $N$}
        \State Compute $\text{IN}(n_i) = \text{OUT}(n_{i-1})$
        \State Compute $\text{OUT}(n_i) = \text{GEN}(n_i) \cup (\text{IN}(n_i) \setminus \text{KILL}(n_i)) \cup \text{FUN}(n_i)$
    \EndFor
    \State \textbf{return} $\{\text{IN}(n_i), \text{OUT}(n_i) \mid \forall i \in \{1, 2, \ldots, N\} \}$
\EndFunction

\State \textbf{do} create a new shared variable $fusion$ that connects to both programs
\State \textbf{let} $FS^A$ $\gets$ \Call{ComputeDataflowSets}{$A$} \Comment{sets of dataflow facts}
\State \textbf{do} identify all dataflows $DF^A$ in $FS^A$ by dataflow propagation
\State \textbf{let} $df^A_i \in DF^A$ be each dataflow in $DF^A$
\State \textbf{let} $w^A_i \gets$ number of variables for each dataflow $df^A_i$
\State \textbf{let} $df^A_{\text{selected}} \gets$ weighted random selection using $\{w^A_1, w^A_2, \ldots, w^A_{|DF^A|}\}$
\State \textbf{let} $V^A \gets$ the set of variables in $df^A_{\text{selected}}$
\State \textbf{let} $v^A \gets$ random variable selection from $V^A$ to be replaced
\State \textbf{do} identify all locations $L^A = \{l_1, l_2, \ldots, l_m\}$ in $A$ where $v^A$ appears
\For{each $l \in L^A$}
    \State \textbf{do} replace $v^A$ with $fusion$ at location $l$ with a probability $p$
\EndFor
\State \textbf{let} $A' \gets A$ \Comment{new program after replacements} 
\State \textbf{do} repeat lines 12-22 for program $B$ and get $B'$
\State \textbf{let} $F \gets A' + B'$ \Comment{with shared variable, concatenate new programs}
\State \textbf{return} $F$ \Comment{finish dataflow fusion, return fused program}
\end{algorithmic}
\end{footnotesize}
\end{algorithm}

The fusion process involves creating a shared variable called \texttt{\$fusion} to connect programs $A$ and $B$. The algorithm identifies all dataflows within program $A$ by propagating among sets of dataflow facts via dataflow analysis. Next, it assigns weights to these dataflows based on the number of variables. The algorithm then randomly selects a dataflow $df^A_{selected}$ from program $A$ using a weighted selection process. Within this selected dataflow, a variable is chosen randomly, and all its locations in program $A$ are identified.

The final steps of the algorithm involve randomly replacing occurrences of the selected variable in program $A$ with the shared variable \texttt{\$fusion} with probability $p$ (we used $p$=0.5), resulting in a modified program $A'$. The same process is repeated for program $B$, yielding a modified program $B'$. The fused program $F$ is then created by concatenating $A'$ and $B'$ with the shared variable, completing the dataflow fusion. The algorithm returns the fused program $F$ as the final output.

We highlight that compared to a naive approach to bridge the last variable of the first test to the first variable of the second test, we introduce more diversity (and randomness) to create many more possible versions of the fused test case which in turn leads to new coverage as shown by the experiments. We introduce the following heuristics: (i) Random dataflow. Each test can have multiple dataflows. We assign different weights to each dataflow and randomly select the fused dataflow based on the weights. (ii) Random variable. Each dataflow can have multiple variables. \name randomly picks one variable as the connection variable to be replaced by an intermediate variable. (iii) Random replacement. When interleaving dataflow, \name randomly picks one or more places of each variable from the source code and replaces them with an intermediate variable that connects to both tests.

The example presented in Figure~\ref{fig:overview} shows the power of dataflow interleaving and the heuristics above. 
Our approach changes the semantics of the test B by replacing the last occurrence of \code{\$gen} (\textie \code{clone \$gen} becomes \code{clone \$fusion}). Note that this particular fusion has maintained the semantics of \code{gen()} using yield. 
% Our approach only replaces the last show up of variable \code{\$gen} in test B, which substitutes the \code{clone()} statement but maintains the semantic of \code{yield} in the \code{gen()} function.
% \ry{with code formatting: 
% Our approach changes the semantics of Test B. Replacing the last occurence of \$gen, i.e. clone \$gen becomes clone \$fusion. Note that particular fusion has maintained the semantics of gen() using yield.
% }
We contrast with a naive dataflow fusion that connects the last variable of test A and the first variable of test B (\textie \code{\$iter} and \code{\$gen}), this does not find the memory error since the semantics is different.
% \ry{updated above}
% we cannot reproduce this memory error due to broken semantics. 
Another possible fusion is to link the final \code{clone} statement with an empty object. We have tested and such a trivial fusion does not trigger the bug.
% found that such easy fusion does not trigger memory errors.

% \yc{TODO: highlight the potential usages in other programming languages. see algo1, not limited in PHP}
% \ry{maybe this TODO can be in the conclusion}

\subsection{Further Strategies}

\label{sec:strategies}

We introduce three helpful strategies along with dataflow fusion when combining seed tests. 

\vspace{+2mm}
\textbf{Test mutation}. In addition to dataflow fusion, \name adds mutations to seed programs before they are analyzed and fused to provide more code semantics. Our approach introduces small differences by replacing expressions and operands. 
% (i) Replace expressions. \name replaces existing expressions with special values like \code{null}, \code{php\_max\_int}, or existing variables. (ii) Replace operands. \name replaces arithmetic, assignment, comparison, and logical operands interchangeably. For example, considering the assignment operands, we replace the common assignment operand \code{=} with the following random value (\code{[`+=',`-=',`*=',`/=',`\%=']}). 
We list some demonstrative test mutations in the following table with examples. They are effective while maintaining valid syntax. Each test mutation is applied at a small rate to introduce extra randomness to the seed programs.

% \vspace{-\topsep}
\begin{table}[h]
\centering
\begin{adjustbox}{width=0.46\textwidth}
\begin{tabular}{llll}
\toprule
\textbf{ID} & \textbf{Expression} & \textbf{Replacements} & \textbf{Example} \\ \hline
01 & arithmetic & other arithmetic operands & \tablecode{\$a = \$b \sout{(\%)}+ 1} \\
02 & assignment & other assignment operands & \tablecode{\$a \sout{(=)}*= \$b + 1} \\
03 & logical & other logical operands & \tablecode{\$a = \$b \sout{(and)}or 1} \\
04 & integer & special values like int\_max & \tablecode{\$a = \$b + \sout{(1)}int\_max} \\
05 & string & special values like null & \tablecode{\$a = \$b.\sout{("a")}null}\\ 
06 & variable & other variables & \tablecode{\$a = \sout{(\$b)}\$c + 1} \\
\bottomrule
\end{tabular}
\end{adjustbox}
% \label{table:mutations}
\end{table}
\vspace{-3mm}

\vspace{+2mm}
\textbf{Interface fuzzing}. Dataflow fusion introduces new, more complex code semantics. \name fuzzes internal PHP  interfaces (\textie existing functions implemented by PHP developers) in the enriched code semantics of fused tests. 
The insight is to test the robustness of function implementations with more complex code semantics, which might not have been tested previously. 
Specifically, \name inserts additional statements both above and below the fused program. 
The inserted statements include a fuzzing function that invokes random PHP functions and collects variables (using \code{get\_defined\_vars()}) from the fused test as the function arguments. For obtaining available PHP function interfaces, \name uses internal functions (\texteg \code{get\_defined\_functions()}, \code{get\_loaded\_extensions()}, and \code{get\_extension\_funcs()}) to dynamically fetch around 1,682 functions (excluding posix-related functions) as the interface candidates. \name uses the \code{ReflectionFunction} PHP class and its method \code{getNumberOfParameters()}, \code{getParameters()}, and \code{getType()} to determine the number and types of function parameters dynamically. \name also inserts various logging statements to record execution for bug reproduction. 

\begin{lstlisting}[language=PHP]
<?php
  class A { public $prop { get {} } }
  class B extends A { }
  get_class_vars('B');
/* UBSan: apply non-zero offset to null pointer */
\end{lstlisting}

The reduced program above shows an unknown segmentation fault \name found via interface fuzzing. We call the function (\textie \code{get\_class\_vars()}) by reusing the variable (reduced) with the value \code{``B''} from the fused code.

\vspace{+2mm}
\textbf{Environment crossover}. \name merges other sections of \tests and inserts additional randomness of PHP configurations. Other significant sections contain the \code{-{}-extension-{}-} and \code{-{}-ini-{}-}, which specifies the execution environments of executing the tests. When fusing dataflow, \name also performs a crossover mutation by merging other sections. Additionally, for \code{-{}-{}ini-{}-} section, the official test cases provide a wide range of options so that \name collects them as a dictionary and applies configurations randomly. Interesting configurations include memory limits, JIT mode, optimization levels, opcache,\footnote{\url{https://www.php.net/manual/en/book.opcache.php}} and script preload. \name fuzzes the PHP interpreter with fused code semantics along with changing environments. This strategy brings extra effectiveness to memory error detection. For example, we detect 3 memory errors related to phpdbg\footnote{\url{https://www.php.net/manual/en/book.phpdbg.php}} by merging the debugging instructions in \code{-{}-phpdbg-{}-} sections.

\section{Implementation}

\label{sec:implementations}

% We developed our automatic fuzzing tool, \name, with over 3,000 lines of Python code. 

% \vspace{+2mm}
\textbf{\name fuzzer}. \name's fuzzer is built upon the official testing script used in the PHP interpreter. This script checks all ``.phpt'' files, parses each section, and invokes the PHP interpreter to gain execution results. However, the script is not designed to handle customized ``.phpt'' files and occasionally terminates the fuzzing process unexpectedly. To address this, we patched the script by commenting out several lines, increasing its tolerance for execution failures, and ensuring our fuzzing completes successfully in parallel.

% \textbf{Host protection}. We try to minimize the impact that \name brings to the host environment. The PHP interpreter supports various system-level functions. In \name, we disable all \href{https://www.php.net/manual/en/book.posix.php}{posix} functions which might be called via random fuzzing of internal interfaces. 

\vspace{+2mm}
\textbf{Bug verifier and PHP program reducer}. We use the reproducing scripts generated by the official script to verify detected memory errors. After verification, we rerun verified memory errors in the normal PHP interpreter without sanitizer to observe their outputs (\textie crash or not). We implemented our PHP program reducer based on the principles of delta debugging~\cite{zeller2002simplifying}. The algorithm systematically comments out specific lines or groups of lines to determine whether the bug oracle (sanitizers) continues to trigger an abort. If the issue persists, those lines are discarded, resulting in a reduced version of the program. This process is repeated iteratively until no smaller reproducer can be found. Our reducer effectively reduces the bug-inducing program (typically reducing it to $\sim 10\%$ of its original size).
% while maintaining efficiency. Additionally, 
Other tools like C-Reduce~\cite{regehr2012test} can be applied to further reduce the program size.

% \textbf{Incremental test corpus}. The key insight behind incremental corpus is to gradually add tests that introduce new code coverage as new seed tests in the test corpus. However, to achieve this, code coverage probing must occur after every execution, considerably increasing execution time; otherwise, we cannot determine which test contributes to higher coverage across multiple runs. We have identified this potential optimization as future work and do not claim it as a contribution or evaluate it further.

\section{Evaluation}

\label{sec:evaluation}

In this section, we answer the following questions to assess various important aspects of \name:

\begin{itemize}[noitemsep,topsep=0.2mm,parsep=0.2mm,partopsep=0pt,leftmargin=*]
\setlength{\itemsep}{1pt}
\setlength{\parsep}{0pt}
\setlength{\parskip}{0pt}
    \item \textbf{New memory errors}. 
    How effective is \name in discovering new memory errors in the PHP interpreter?
    \item \textbf{Improved effectiveness}. 
     To what extent does \name improve the effectiveness compared to (i) the official test suite and (ii) test concatenation? 
    \item \textbf{Comparison with existing approaches}. What is the improvement in fuzzing the PHP interpreter with the state-of-the-art fuzzing approaches?
    \item \textbf{Ablation study}. What is the impact of each strategy of \name on contributing to the overall effectiveness?
\end{itemize}

\vspace{+2mm}
\textbf{PHP version and compilation}. In Section~\ref{sec:bugs}, as well as during daily fuzzing, we compiled the latest commit of PHP interpreter using clang-15,\footnote{
We chose Clang given it has greater and more accurate sanitizer support, also shown by evaluation of memory error defenses~\cite{jiang2022recipe}.
% We chose clang due to its higher accuracy with sanitizers, as supported by existing research~\cite{jiang2022recipe}.
} with debug symbols, address sanitizers, and undefined behavior sanitizers enabled. In other sections (\ref{sec:improvement}, \ref{sec:comparison}, and \ref{sec:ablation}), we performed evaluations on a specific commit (\textie v8.3.3, 3a832a2aad405466c24a5e8e5798cf9de14fda14) of the PHP interpreter. For Sections~\ref{sec:improvement} and~\ref{sec:ablation}, we compiled the PHP interpreter using Ubuntu 22.04’s default CC (gcc-11.4.0) for better compatibility, with debug symbols, sanitizers, and gcov support enabled. In Section~\ref{sec:comparison}, we instead used afl-gcc for comparison with existing work, while similarly enabling debug symbols, sanitizers, and gcov support.

\vspace{+2mm}
\textbf{Evaluation metrics}. We explain two evaluation metrics we used in this section to measure the effectiveness of \name and related approaches. (i) Code coverage. Code coverage is a widely used metric for evaluating the effectiveness of fuzzing approaches~\cite{rigger2020finding, Rigger2020PQS}. We collect code coverage using 
% publicly available
gcovr~\cite{gcovr}. We
follow the suggested gcovr from the official Makefile to measure the overall code coverage. We only report line code coverage in this work and omit the other coverage metrics because they follow a similar trend. (ii) Number of unique crash sites. We define the crash site as the unique file path of the PHP interpreter and line number from the sanitizer abort, which we use to approximate the number of detected bugs. For example, the motivating example in Listing~\ref{lst:motivate} has the following sanitizer abort ``AddressSanitizer: heap-use-after-free \textit{/php-src/ext/dom/php\_dom.c:311}'' with the crash site being highlighted in italics. After a period of fuzzing, the number of unique crash sites can be used to de-duplicate similar aborts, hence demonstrating the effectiveness of various fuzzing approaches. Based on our experience, most previously unknown crash sites can be reported as bugs to the PHP interpreter and are likely to be fixed.

\vspace{+2mm}
\textbf{Experimental infrastructure}. All experiments were conducted on an AMD EPYC 7763 processor with 64 physical cores and 128 logical cores, clocked at 2.45 GHz. The test machine ran Ubuntu 22.04 and was equipped with 512 GB of RAM. 
By default, \name uses a modest computing resource, 32 CPU cores with up to 32 GB of RAM, capable of finding various crash sites within 24 hours.

\subsection{Discovering Unknown Memory Errors}

\label{sec:bugs}

To discover unknown memory errors in the PHP interpreter, we intermittently tested the latest version of the PHP interpreter, compiled from the php-src repository, over a six-month period. This approach aligns with standard methodologies for evaluating the effectiveness of automatic bug-finding tools~\cite{Rigger2020PQS, kamm2022testing}. Next, we reduced the merged test cases and verified whether the issues had already been reported on issue trackers to avoid duplicate bug reports. Bug-inducing test cases generated by \name are often complex; we automatically simplified them to smaller, bug-inducing versions using delta debugging~\cite{zeller2002simplifying}. Below, we illustrate memory errors using these reduced PHP programs.

\vspace{+2mm}
\textbf{Results}. 
% \ry{Table 1 is less than all results, also cannot list all as too many}
Table~\ref{table:bugs} 
% lists all reported memory errors 
shows information about the first 50 bugs found by \name.
% found by our approach and verified by humans. 
% \ry{YC: see if any of the integer overflow ones can lead to memory errors, otherwise may have to remove the CWE-190 (overflow) ones.}
The column \textit{Bug Type} shows the corresponding Common Weakness Enumeration (CWE) type of bugs\footnote{The bugs found by \name covered the following CWEs: CWE-121 Stack-based Buffer Overflow, CWE-122 Heap-based Buffer Overflow, CWE-124 Buffer Underwrite, CWE-190 Integer Overflow or Wraparound, CWE-401 Missing Release of Memory after Effective Lifetime, CWE-416 Use After Free, CWE-457 Use of Uninitialized Variable, CWE-476 NULL Pointer Dereference, CWE-824 Access of Uninitialized Pointer, CWE-825 Expired Pointer Dereference} to concisely present their root causes. All reported memory errors were identified with the sanitizer oracles, and the \textit{Crash} column shows bugs that could crash the program even without sanitizers enabled, highlighting cases with more severe impacts. The \textit{Bug Location} provides the accurate bug locations 
% (\textie links to official code) 
of all fixed bugs. The \textit{Issue ID} column provides the official GitHub issue number of the corresponding bug. The \textit{Status} shows the bug status. 
We classified the bug status into the following disjoint categories: 
\begin{itemize}[noitemsep,topsep=0.2mm,parsep=0.2mm,partopsep=0pt,leftmargin=*]
\setlength{\itemsep}{1pt}
\setlength{\parsep}{0pt}
\setlength{\parskip}{0pt}
    \item[-] \textit{Pending} (\textbf{\texttt{Pd}}) bugs refer to the bugs submitted but awaiting further investigation to confirm the root cause.
    \item[-] \textit{Expected} (\textbf{\texttt{Ep}}) bugs refer to the bugs been confirmed but developers believe they are expected behaviors.
    \item[-] \textit{Confirmed} (\textbf{\texttt{Cf}}) bugs refer to the bugs that have been confirmed but have not been fixed.
    % \item[-] \textit{Duplicated} (\textbf{\texttt{Dp}}) bugs refer to the bugs that have been confirmed but are regarded as duplicates of existing issues.
    \item[-] \textit{Fixed} (\textbf{\texttt{Fx}}) bugs refer to the bugs that have been confirmed and patched by the developers.
\end{itemize}

\begin{table}[t!]
\centering
\caption{First 50 Bugs Found by \name}
\begin{adjustbox}{width=0.48\textwidth}
\begin{tabular}{clcllll} 
\toprule
\textbf{ID} & \textbf{Bug Type} & \textbf{Crash} & \textbf{Bug Location} & \textbf{Issue ID} & \textbf{Status} & \textbf{Fixes} \\ \hline
01 & CWE-825 & \checkmark &  \href{https://github.com/php/php-src/blob/master/ext/session/session.c}{session.c} & \issuelink{13680} & \textbf{\texttt{Fx}} & \tablefix{28}{2} \\
02 & CWE-476
 & \checkmark &  \href{https://github.com/php/php-src/blob/master/sapi/phpdbg/phpdbg_watch.c}{phpdbg\_watch.c}
 & \issuelink{13681} & \textbf{\texttt{Fx}} & \tablefix{51}{6} \\
03 & CWE-476 & \checkmark & \href{https://github.com/php/php-src/blob/master/ext/spl/spl_directory.c}{spl\_directory.c} & \issuelink{13685} & \textbf{\texttt{Fx}} &\tablefix{80}{18} \\
04 & CWE-121 & - & - & \issuelink{13768} & \textbf{\texttt{Ep}}\\
05 & CWE-476 & \checkmark & \href{https://github.com/php/php-src/blob/master/sapi/phpdbg/phpdbg_frame.c}{phpdbg\_frame.c} & \issuelink{13827} & \textbf{\texttt{Fx}} & \tablefix{41}{2} \\
06 & CWE-476 & \checkmark & \href{https://github.com/php/php-src/blob/master/ext/phar/phar.c}{phar.c} & \issuelink{13833} & \textbf{\texttt{Fx}} & \tablefix{68}{20} \\
07 & CWE-476 & \checkmark & \href{https://github.com/php/php-src/blob/master/ext/opcache/jit/zend_jit.c}{zend\_jit.c}, .. & \issuelink{13834} & \textbf{\texttt{Fx}} & \tablefix{29}{24} \\
08 & CWE-476 & \checkmark & \href{https://github.com/php/php-src/blob/master/ext/phar/stream.c}{stream.c} & \issuelink{13836} & \textbf{\texttt{Fx}} & \tablefix{36}{2} \\
09 & CWE-476 & \checkmark & \href{https://github.com/php/php-src/blob/master/ext/session/mod_user_class.c}{mod\_user\_class.c} & \issuelink{13856} & \textbf{\texttt{Fx}} & \tablefix{23}{2} \\
10 & CWE-190 & - & - & \issuelink{13881} & \textbf{\texttt{Pd}}\\
11 & CWE-124 & - & - & \issuelink{13903} & \textbf{\texttt{Ep}}\\
12 & CWE-476 & \checkmark & \href{https://github.com/php/php-src/blob/master/main/main.c}{main.c} & \issuelink{13931} & \textbf{\texttt{Fx}} & \tablefix{52}{0} \\
13 & CWE-122 & \checkmark & \href{https://github.com/php/php-src/blob/master/ext/pdo_sqlite/sqlite_driver.c}{sqlite\_driver.c} & \issuelink{13984} & \textbf{\texttt{Fx}} & \tablefix{19}{1} \\
14 & CWE-457 & \checkmark & \href{https://github.com/php/php-src/blob/master/ext/pdo_sqlite/sqlite_driver.c}{sqlite\_driver.c} & \issuelink{13998} & \textbf{\texttt{Fx}} & \tablefix{25}{1} \\
15 & CWE-401 & - & \href{https://github.com/php/php-src/blob/master/ext/xml/compat.c}{compat.c} & \issuelink{14044} & \textbf{\texttt{Fx}} & \tablefix{35}{5} \\
16 & CWE-190 & - & - & \issuelink{14075} & \textbf{\texttt{Pd}}\\
17 & CWE-476 & \checkmark & - & \issuelink{14082} & \textbf{\texttt{Cf}}\\
18 & CWE-824 & \checkmark & \href{https://github.com/php/php-src/blob/master/ext/xml/xml.c}{xml.c} & \issuelink{14124} & \textbf{\texttt{Fx}} & \tablefix{28}{0} \\
19 & CWE-121 & \checkmark & - & \issuelink{14164} & \textbf{\texttt{Ep}}\\
20 & CWE-476 & \checkmark & \href{https://github.com/php/php-src/blob/master/ext/spl/spl_iterators.c}{spl\_iterators.c} & \issuelink{14290} & \textbf{\texttt{Fx}} & \tablefix{24}{3} \\
21 & CWE-401 & - & \href{https://github.com/php/php-src/blob/master/ext/dom/document.c}{document.c} & \issuelink{14343} & \textbf{\texttt{Fx}} & \tablefix{24}{1} \\
22 & CWE-121 & \checkmark & \href{https://github.com/php/php-src/blob/master/Zend/zend_compile.c}{zend\_compile.c}, .. & \issuelink{14361} & \textbf{\texttt{Fx}} & \tablefix{54}{8} \\
23 & CWE-476 & \checkmark & \href{https://github.com/php/php-src/blob/master/ext/phar/zip.c}{zip.c} & \issuelink{14603} & \textbf{\texttt{Fx}} & \tablefix{1}{1} \\
24 & CWE-476 & \checkmark & \href{https://github.com/php/php-src/blob/master/ext/simplexml/simplexml.c}{simplexml.c} & \issuelink{14638} & \textbf{\texttt{Fx}} & \tablefix{57}{24} \\
25 & CWE-476 & \checkmark & \href{https://github.com/php/php-src/blob/master/ext/spl/spl_observer.c}{spl\_observer.c} & \issuelink{14639} & \textbf{\texttt{Fx}} & \tablefix{29}{4} \\
26 & CWE-824 & \checkmark & \href{https://github.com/php/php-src/blob/master/ext/standard/basic_functions.c}{basic\_functions.c} & \issuelink{14643} & \textbf{\texttt{Fx}} & \tablefix{5}{2} \\
27 & CWE-824 & \checkmark & \href{https://github.com/php/php-src/blob/master/ext/dom/php_dom.c}{php\_dom.c} & \issuelink{14652} & \textbf{\texttt{Fx}} & \tablefix{21}{1} \\
28 & CWE-825 & \checkmark & \href{https://github.com/php/php-src/blob/master/ext/spl/spl_directory.c}{spl\_directory.c} & \issuelink{14687} & \textbf{\texttt{Fx}} & \tablefix{42}{1} \\
29 & CWE-825 & \checkmark & \href{https://github.com/php/php-src/blob/master/ext/libxml/libxml.c}{libxml.c} & \issuelink{14698} & \textbf{\texttt{Fx}} & \tablefix{29}{3} \\
30 & CWE-190 & - & - & \issuelink{14709} & \textbf{\texttt{Pd}} & \\
31 & CWE-476 & \checkmark & \href{https://github.com/php/php-src/blob/master/Zend/zend_execute.c}{zend\_execute.c}, .. & \issuelink{14712} & \textbf{\texttt{Fx}} & \tablefix{32}{1} \\
32 & CWE-190 & - & - & \issuelink{14732} & \textbf{\texttt{Pd}} & \\
33 & CWE-825 & \checkmark & \href{https://github.com/php/php-src/blob/master/Zend/zend_interfaces.c}{zend\_interfaces.c} & \issuelink{14741} & \textbf{\texttt{Fx}} & \tablefix{18}{0} \\
34 & CWE-190 & - & \href{https://github.com/php/php-src/blob/master/ext/standard/basic_functions.c}{basic\_functions.c} & \issuelink{14774} & \textbf{\texttt{Fx}} & \tablefix{29}{0} \\
35 & CWE-190 & - & \href{https://github.com/php/php-src/blob/master/ext/standard/array.c}{array.c} & \issuelink{14775} & \textbf{\texttt{Fx}} & \tablefix{16}{0} \\
36 & CWE-190 & - & \href{https://github.com/php/php-src/blob/master/ext/standard/file.h}{file.h} & \issuelink{14780} & \textbf{\texttt{Fx}} & \tablefix{62}{7} \\
37 & CWE-476 & \checkmark & \href{https://github.com/php/php-src/blob/master/main/output.c}{output.c} & \issuelink{14808} & \textbf{\texttt{Fx}} & \tablefix{18}{1} \\
38 & CWE-476 & \checkmark & \href{https://github.com/php/php-src/blob/master/ext/dom/text.c}{text.c} & \issuelink{15137} & \textbf{\texttt{Fx}} & \tablefix{14}{1} \\
39 & CWE-190 & - & - & \issuelink{15150} & \textbf{\texttt{Pd}} & \\
40 & CWE-121 & \checkmark & - & \issuelink{15168} & \textbf{\texttt{Cf}} & \\
41 & CWE-121 & \checkmark & - & \issuelink{15169} & \textbf{\texttt{Cf}}\\
42 & CWE-476 & \checkmark & \href{https://github.com/php/php-src/blob/master/ext/standard/url_scanner_ex.re}{url\_scanner\_ex.re} & \issuelink{15179} & \textbf{\texttt{Fx}} & \tablefix{24}{1} \\
43 & CWE-476 & \checkmark & \href{https://github.com/php/php-src/blob/master/main/output.c}{output.c} & \issuelink{15181} & \textbf{\texttt{Fx}} & \tablefix{19}{0} \\
44 & CWE-825 & \checkmark & - & \issuelink{15187} & \textbf{\texttt{Cf}} & \\
45 & CWE-825 & \checkmark & \href{https://github.com/php/php-src/blob/master/ext/dom/php_dom.c}{php\_dom.c} & \issuelink{15192} & \textbf{\texttt{Fx}} & \tablefix{77}{1} \\
46 & CWE-824 & \checkmark & \href{https://github.com/php/php-src/blob/master/sapi/phpdbg/phpdbg_bp.c}{phpdbg\_bp.c} & \issuelink{15208} & \textbf{\texttt{Fx}} & \tablefix{57}{0} \\
47 & CWE-416 & \checkmark & \href{https://github.com/php/php-src/blob/master/sapi/phpdbg/phpdbg.h}{phpdbg.h}, .. & \issuelink{15210} & \textbf{\texttt{Fx}} & \tablefix{75}{3} \\
48 & CWE-122 & \checkmark & \href{https://github.com/php/php-src/blob/master/sapi/phpdbg/phpdbg_info.c}{phpdbg\_info.c} & \issuelink{15210} & \textbf{\texttt{Fx}} & \tablefix{34}{6} \\
49 & CWE-416 & \checkmark & \href{https://github.com/php/php-src/blob/master/ext/dom/nodelist.c}{nodelist.c} & \issuelink{15143} & \textbf{\texttt{Fx}} & \tablefix{29}{2} \\
50 & CWE-416 & \checkmark & \href{https://github.com/php/php-src/blob/master/ext/pcre/php_pcre.c}{php\_pcre.c}, .. & \issuelink{15205} & \textbf{\texttt{Fx}} & \tablefix{58}{47} \\
\bottomrule
\end{tabular}
\end{adjustbox}
\label{table:bugs}
\end{table}

% \begin{figure}[t]
% \begin{lstlisting}[escapeinside={(*}{*)}, language=PHP, label=lst:20years, caption=Heap Use-after-free in DOM Extension]
% <?php
% $dom = new DOMDocument;
% $dom->loadXML(<<<XML
% <!DOCTYPE foo [
% <!ENTITY foo1 "bar1">
% ]>
% <foo>&foo1;</foo>
% XML);
% $ref = $dom->documentElement->firstChild;
% $nodes = $ref->childNodes;
% $dom->removeChild($dom->doctype);
% $script1_dataflow = $nodes;
% foreach($script1_dataflow as $str) {}
% /* AddressSanitizer: heap use-after-free .. */
% \end{lstlisting}
% \end{figure}

%28+51+80+41+68+29+36+23+52+19+35+28+24+24+54+1+57+29+5+21+42+29+32+18+29+16+62+18+14+24+19+77+57+75+34+29+58=1338
%2+6+18+2+20+24+2+2+1+1+5+3+1+8+1+24+4+2+1+1+3+1+7+1+1+1+1+3+6+2+47=201
% ['session.c','phpdbg_watch.c','sql_directory.c']

\vspace{+2mm}
\textbf{Bug diversity.} In total, we identified \totalbug previously unknown bugs.
% \footnote{Bug number is still increasing as we continuously find unknown errors every week when we finalize this paper.} 
Of these, \pendbug are still pending triage, \expectbug were expected, \confbug have been confirmed, \dupbug were marked as duplicates, and \fixbug have already been fixed. All \totalbug errors were detected by sanitizers, and \crashbug of them caused the PHP interpreter to crash when compiled normally without sanitizers. We present the first 50 bugs listed in Table~\ref{table:bugs}, which covers nearly all common types of memory errors. This also demonstrates that \name can significantly enhance the security of the PHP interpreter finding many different errors across different modules in PHP. From the bugs in Table~\ref{table:bugs}, \name identified 8 instances of out-of-bounds errors (\texteg stack/heap buffer overflows/underflows) and 3 use-after-free bugs. Additionally, \name detected 18 null-pointer dereference bugs, 1 use-of-uninitialized-variable bug, 2 expired-pointer dereference bugs, and 6 access-of-uninitialized-pointer bugs, all of which involve ``bad''
% illegal
% \ry{technically memory error is NOT illegal, it is UNDEF which we call bad}
pointer access that can cause the PHP interpreter to crash unexpectedly. We also detected several non-crash bugs, including 3 memory leaks and 9 signed integer overflows.
% \ry{YC: numbers dont add up to 50, check}

We found bugs affecting diverse functionalities within the PHP interpreter, as illustrated in the locations of fixed bugs, which are partly shown in Table~\ref{table:bugs}. In total, patches generated by \name have impacted over 80 individual files, including critical source files (\texteg zend\_*.c, main.c). Specifically, the developers changed over 5k lines of code as a direct response to our reports for security improvements in the official PHP interpreter repository. The bugs found by \name mostly have existed in the PHP interpreter from several months to years. One use-after-free bug (the motivating example shown in Listing~\ref{lst:motivate}) has existed for more than 20 years, which developers described as ``\textit{an ancient bug}''.

Next, we highlight notable bugs found by \name, describing their root causes and security impacts based on our analysis and developers' feedback.

% \textbf{Case study 1: (>20 years) (high severity) heap use-after-free in DOM extension.} The PHP \href{https://www.php.net/manual/en/book.dom.php}{DOM} allows operations on XML and HTML documents, which is the key component supporting the websites. \name detected a heap use-after-free bug that has existed for more than 20 years, as shown in Listing~\ref{lst:motivate}. This bug arises due to removing an entity declaration from the document, but then
% entity references will keep pointing to that stale declaration in the \code{foreach} iteration. \name detects this bug by dataflow fusion via connecting variables \code{\$nodes} and \code{\$values}. 

\begin{figure}[t]
\begin{lstlisting}[escapeinside={(*}{*)}, language=PHP, label=lst:UAF, caption=Heap use-after-free in PCRE, aboveskip=-3pt, belowskip=0pt]
<?php
  $array = new ArrayIterator(..);
  $regex = new RegexIterator($array, '/Array/');
  foreach ($regex as $match) { }
  $fusion = $regex;
/* AddressSanitizer: heap-use-after-free */
\end{lstlisting}
\end{figure}

\begin{figure}[t]
\begin{lstlisting}[escapeinside={(*}{*)}, language=PHP, label=lst:sqlite_bug, caption=Heap overflow in SQLite extension, aboveskip=-3pt, belowskip=-5pt]
<?php
  $dbfile = $GLOBALS[array_rand($GLOBALS)];
  $db = new PDO('sqlite:'.$dbfile, null, null, ..);
/* AddressSanitizer: heap-buffer-overflow on .. */
\end{lstlisting}
\end{figure}

\begin{figure}[t]
\begin{lstlisting}[escapeinside={(*}{*)}, language=PHP, label=lst:zend_bug1, caption=Null pointer dereference in Zend compiler, aboveskip=0pt, belowskip=0pt]
<?php
  register_shutdown_function(function() {
    var_dump(eval("return 1+3;")); });
  eval(<<<EVAL
    function f(){ try { break; } finally {}} f(); 
    EVAL);
/* UBSan: applying zero offset to null pointer */
\end{lstlisting}
\end{figure}

% /* test.php */
% <?php
%   Foo::test();
% /* preload.inc */
\begin{figure}[t]
\begin{lstlisting}[escapeinside={(*}{*)}, language=PHP, label=lst:zend_bug2, caption=Segmentation fault in JIT and OPcache, aboveskip=0pt, belowskip=0pt]
<?php
  class Foo { public static function test() { 
      static $i = 0; var_dump(++$i); } }
  Foo::test();
/* AddressSanitizer: SEGV on unknown address */
\end{lstlisting}
\end{figure}

% \vspace{-5mm}
\begin{figure}[t]
\begin{lstlisting}[escapeinside={(*}{*)}, language=PHP, label=lst:session_bug, caption=Segmentation fault in session extension, aboveskip=0pt, belowskip=0pt]
<?php
  ob_start();
  ini_set("session.serialize_handler", ..);
  session_start();
  $result1 = session_decode('foo|s:3:"bar";');
  class Test extends DateTime {
    public static function createFromFormat(
    $format, $datetime, $timezone = null): Wrong{} 
  }
/* AddressSanitizer: SEGV on unknown address */
\end{lstlisting}
\end{figure}

\vspace{+1mm}
\textbf{Bug analysis 1: heap use-after-free with PCRE.} The PHP Perl Compatible Regular Expressions (PCRE) module facilitates pattern matching in PHP scripts using functions like \code{preg\_match()}. A heap use-after-free vulnerability, illustrated in Listing~\ref{lst:UAF}, arises due to the premature shutdown of the PCRE module before all live objects are destroyed. Consequently, when the Standard PHP Library (SPL) attempts to clean up a regular expression object, it operates on freed memory, leading to a use-after-free error.

\name discovered this bug by reassigning the regular expression variable after completing all regular expression operations, guided by dataflow fusion. The original query included a series of standard regular expression statements along with another script containing typical assignments. As described in Section~\ref{sec:algorithm}, \name randomly replaces variables to create bridging connections for dataflow fusion, potentially linking regular expression variables with subsequent unrelated assignments, which then triggered this bug.

\vspace{+1mm}
\textbf{Bug analysis 2: heap overflow in PHP SQLite module.} 
SQLite is a widely used lightweight database system embedded in the PHP interpreter as an in-memory storage solution. Listing~\ref{lst:sqlite_bug} illustrates a heap overflow error in the SQLite module within the PHP interpreter. This bug-triggering program first assigns a database variable as the SQLite file and then creates a new SQLite object. However, because the buffer size is not checked before \textit{memcmp}, the allocated buffer exceeds the expected size in the heap, triggering an alert from ASan. \name
% initially
found this issue by interleaving unrelated variables with SQLite statements.
% which was promptly acknowledged and resolved.

\vspace{+1mm}
\textbf{Bug analysis 3: null pointer dereference in the Zend compiler.} The Zend engine contains a null pointer dereference vulnerability, as shown in Listing~\ref{lst:zend_bug1}. This bug first causes the compiler to terminate due to a fatal error, leaving its data structures with stale values. During the next compilation, elements from the previous stack are incorrectly reused, leading to a segmentation fault because wrong instructions are emitted due to the stale data. 

\vspace{+1mm}
\textbf{Bug analysis 4: segmentation fault in JIT and opcache.} The PHP interpreter supports just-in-time compilation through its JIT compiler and accelerates execution using opcache. Listing~\ref{lst:zend_bug2} presents a related issue that causes the Zend compiler to crash. This issue was identified in a specific environment with an opcache preload configuration. Developers noted that ``\textit{the issue arises because caller\_info, callee\_info, and possibly call\_map are allocated in the arena but are not reset before being used by the next request}.''

\vspace{+1mm}
\textbf{Bug analysis 5: segmentation fault in the session module.} The session module in the PHP interpreter preserves data across visits by assigning each visitor a unique session ID. This support allows data storage between requests using the \code{\$\_SESSION} superglobal array. Listing~\ref{lst:session_bug} gives a segmentation fault found in the session module. This fault arises from changes made to the session decode process to prevent writing incomplete sessions. According to the developers, ``\textit{it is illegal to return from a bailout because that does not restore the original bailout data}'', the fix makes the termination outside of the original data.

\vspace{+1mm}
\textbf{Discussion---false alarms.} 
We observed three false alarms during our experiment that are linked to the expected (\textbf{\texttt{Ep}}) bugs in Table~\ref{table:bugs}. There are two reasons for these false alarms: (i) \name relies on sanitizers as the bug oracle. However, it is known that sanitizers can have incorrect results~\cite{li2024ubfuzz}.
% \ry{YC: it is unclear if the explanation below is in their testing code as they may have used ASAN APIs, so don't want to imply this one is ASAN's fault}
% may generate incorrect alerts.
% However, it is known that sanitizers can have incorrect results~\cite{li2024ubfuzz}. 
% \ry{Maybe should rework the sentence and remove the ref which is not about FP} \yc{it is fine for me, the point is sanitizers can go wrong}
% \ry{reason is probablu not right, the developer reason is that they are using ASAN APIs but have a bug - check the code for the ASAN APIs}
% The explanation by the developers for one expected bug is:
One of the expected bugs is explained by developers as 
``{\it This is an ASAN false positive $\ldots$ gives us a memory region that is adjacent to the region where the fiber stack used to be, but that still has the old shadow memory, and so we get a false ASAN warning}''. (ii) Another cause of false alarms is that the sanitizer may abort before the overflow handler is triggered. The PHP interpreter has a conservative stack limit, which can sometimes be reached after the sanitizer issues an alert. This results in false alarms for the remaining two expected bugs.
% \ry{YC: we already discuss we dont use expect so this may be confusing. Fix}

% \ry{YC: need to say it is due to specific use of thread memory which is
% deallocated but not managed by ASan so not unpoisoned - in footnote. Normally ASAN has NO FP. the paper is mainly about FN and does not find any FP!.}
% \ry{YC: I will add from the github the reason but fix the above}
% \yc{I do not think explaining in detail will help. the only thing reviewers need to know is that sanitizers can be incorrect. no need to give reasons}

% \textbf{Discussion---bugs exploitability.} Although most of the memory errors discovered by \name are memory errors, such as buffer overflows and null pointer dereferences, advancing to the next step of exploitation is challenging. Exploitation typically requires complex, hacky techniques that are beyond the scope of this paper. Future work could integrate a series of Automatic Exploitation Generation (AEG) to investigate these bugs further or potentially increase their severity.

\subsection{Improvement on Official Test Suite}

\label{sec:improvement}

One way to evaluate the effectiveness of \name is by assessing how well it improves upon the official test suite in the PHP interpreter and the naive approach of test concatenations. To achieve this, we manually analyzed the first 50 bugs identified by \name, as detailed in Table~\ref{table:bugs}. Our findings reveal that only 3 bugs can be reproduced using test concatenation, and none can be reproduced using the official test suite, highlighting the remarkable effectiveness of our new approach.

Next, we performed a 24-hour fuzzing run, as recommended in previous research~\cite{klees2018evaluating}, under three different setups: (a) the official test suite, (b) concatenations of official test cases, and (c) dataflow fusion. Note we merge other sections to ensure a fairer comparison when evaluating the test concatenation. The enhanced effectiveness of \name is demonstrated through two key outcomes from the 24-hour fuzzing: (i) \name discovers more unique crash sites that the official test suite and the test concatenation cannot detect, and (ii) \name achieves higher code coverage compared to the official test suite and the test concatenation.

\begin{figure}[b]
    \centering
    \includegraphics[scale=0.48]{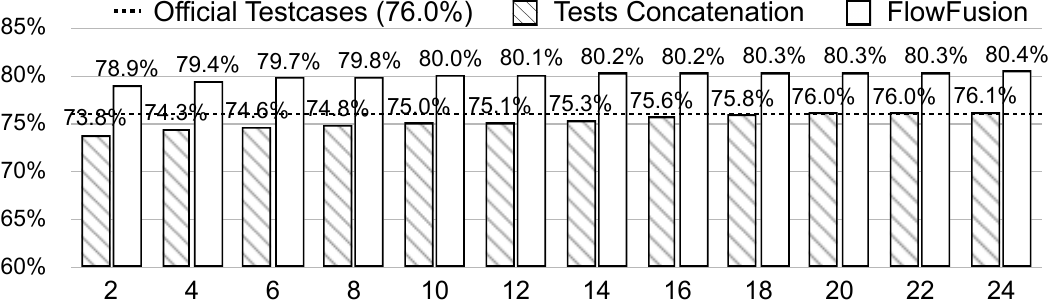}
    \caption{Code Coverage over Test Suite and Concatenation}
    \label{fig:coverage}
\end{figure}

% \begin{figure}[t]

% \end{figure}
\vspace{+2mm}
\textbf{More unique crash sites}. \name is effective in discovering unique crash sites in the PHP interpreter. We track the number of crash sites of three approaches. Our evaluation shows that \name can uncover over 20 unique crash sites on average, whereas the official test cases detect none, and test concatenation detects only 2. We expect the official test cases to only rarely detect these crash sites, as they are tested daily in automated continuous integration. Test concatenation can reveal some crash sites by merging other non-program sections, which creates different execution environments. 

\begin{lstlisting}[escapeinside={(*}{*)}, language=PHP, label=lst:zend_alloc]
<?php
  class Test { public string $prop {
    set => strtoupper($value); } }
  $test = new Test();
  var_dump($test);
  foreach ($test as $longVal) {}
/* AddressSanitizer: SEGV on unknown address */
\end{lstlisting}

The example test above illustrates where the memory error found by \name is missed by both the official test suite and test concatenation. This error occurs in the Zend allocator and is triggered when a non-related class object from one seed test is assigned to a \code{foreach} statement in another test. The official test cases lacked a test to check such code semantics while test concatenation fails to establish connections between merged tests. Thus both approaches fail to find this bug.

% We assess the code coverage of \name compared to the official test suite on the PHP interpreter over 24 hours, as recommended in previous research~\cite{klees2018evaluating}. 

% We evaluate \name against two different settings (i) the official test suite (\textie we keep running the official test cases during the 24 hours) (ii) tests concatenation (\textie we randomly combine two tests from the official test suite during the 24 hours) and (iii) the continuous 24 hours fuzzing by \name to compare its improvement.

\vspace{+2mm}
\textbf{Higher code coverage}. Figure~\ref{fig:coverage} illustrates that \name outperforms both the official test suite and the concatenation tests in terms of code coverage. 
% \ry{YC: Fig 2 - is the Official test coverage 76.0\% or 76.1\% needs to be consistent. Which line is the 76.1?}
Consistent with previous observations, the well-maintained official test cases consistently achieve a high code coverage of 76\% over 24 hours. This result, while impressive and surpassing existing grammar-based fuzzers, remains static due to the limited size of the test suite.

Test concatenation, on the other hand, begins with lower coverage than the official test suite, likely due to compatibility issues or syntax errors introduced during the merging process (\texteg unresolved namespace declarations), which cause some tests to fail. However, the coverage for test concatenation gradually increases over 24 hours, eventually surpassing that of the official test suite. In general, test concatenation demonstrates similar effectiveness to the official test suite.
% This trend supports our hypothesis that while the native approach of test concatenation does introduce some new code semantics, its potential for increasing coverage is limited.

\name, in contrast, achieves 80.4\% with 4.3\% higher code coverage than the official test suite after 24 hours, with coverage continuing to grow. \name explores a vast space of generated test cases to uncover more memory errors. This approach sacrifices some short-term efficiency for long-term gains in code coverage. Notably, \name continues to increase coverage beyond the 24-hour mark. For example, \name reaches 82.1\% code coverage after 7 days, while the test suite's coverage remains unchanged.

% \ry{Not sure if there is a point that the code coverage achieved is very good but need refs for that ... Google coverage seems to be around the ball park of 80\%}

% \textbf{Bug issue trend}. We next explore the bug issue related to memory errors in the official PHP bug tracker before and after we applied \name to detect bugs in the PHP interpreter. 

\begin{figure}[b]
    \centering
    \includegraphics[scale=0.44]{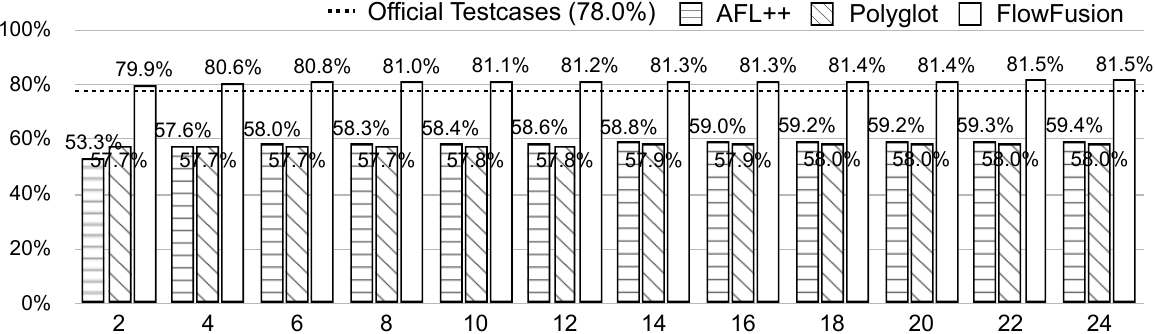}
    \caption{Coverage Comparison against AFL++ and Polyglot}
    \label{fig:polyglot}
\end{figure}

\vspace{-2mm}
\subsection{Comparison with Existing Approaches}

\label{sec:comparison}

We assess the effectiveness of \name in identifying memory errors in the PHP interpreter compared to existing methodologies. 
We included AFL++~\cite{fioraldi2020afl++} since it is often considered the most effective general-purpose fuzzer. However, given that the PHP interpreter expects highly structured input, we also consider grammar-based fuzzing approaches, which we would expect to generate more valid inputs. To the best of our knowledge, no existing approach specifically focuses on PHP interpreter fuzzing.
% in top-tier conferences.
% \ry{cannot say top tier}
We thus consider the following fuzzers designed for general programming languages. Polyglot~\cite{chen2021one}, which incorporates grammar guidance and semantic validation, has uncovered over 30 memory errors in the PHP interpreter, making it the most effective existing approach. % Therefore, we also consider Polyglot to be another baseline in our evaluation.
% \ry{repetition with baseline below, deleted}
Among other grammar-based approaches previously used to detect memory errors in the PHP interpreter, Langfuzz\cite{holler2012fuzzing} is not publicly accessible. Nautilus~\cite{aschermann2019nautilus} and Gramatron~\cite{srivastava2021gramatron} found 3 and 4 memory errors in the PHP interpreter, respectively, indicating their limited effectiveness. Therefore, we select the most effective approach Polyglot as the state-of-the-art evaluation baseline.

To conduct a fair comparison, we extract all PHP programs from the official test suite and make them the seed inputs for AFL++ and Polyglot. We perform a 24-hour fuzzing evaluation by running AFL++, Polyglot, and \name under the same environment (limited to 16 cores for each approach) to compare their code coverage---the direct metric associated with fuzzing effectiveness. We compile three copies of the PHP interpreter using the same AFL-specific compiler\footnote{Coverage results differ from Section~\ref{sec:improvement} as we change to AFL-clang compiler.} and compiler options (with gcov option and sanitizers).

\vspace{+2mm}
\textbf{Higher code coverage}. Figure~\ref{fig:polyglot} shows the results of a 24-hour fuzzing experiment, where \name significantly outperforms both AFL++ and Polyglot. Notably, while using all official test cases as seed programs, AFL++ and Polyglot achieve code coverages of around 60\%, which falls short compared to the 78\% coverage achieved by the official test suite alone. This discrepancy arises due to two main reasons. First, AFL++ and Polyglot are not PHP-specific fuzzers, and face difficulties in fuzzing various sections of the test cases, such as the \code{-{}-ini-{}-} and \code{-{}-extensions-{}-} sections, as illustrated in Listing~\ref{lst:phpt}. 
% \ry{expand or shorten previous}
Consequently, they often fail to dynamically meet the additional module requirements of the generated programs and to mutate the execution environments specified in these configuration sections. This limitation results in inadequate testing of core PHP interpreter features, such as JIT compilation. Second, as the seed programs are already diverse in their semantics and syntactically correct, the room for improvement using grammar guidance and semantic validation in AFL++ and Polyglot is limited. In contrast, \name enhances the official test suite by generating more complex code semantics through dataflow fusion, thereby increasing its semantic diversity.

% \begin{figure}[t]
% \begin{lstlisting}[escapeinside={(*}{*)}, language=PHP, label=lst:5.3, caption=Segmentation fault in PHP URL scanner]
% <?php
%   for ($i=0; $i<=$mb; $i++) {
%     $var.= str_repeat('a',1*1024*1024);
%   }
%   $x=$var;
%   output_add_rewrite_var($x,$x);
% \end{lstlisting}
% \end{figure}

\vspace{+2mm}
\textbf{More memory errors}. We also counted the number of memory errors detected by these three approaches over a 24-hour period. AFL++ and Polyglot detected 49 and 21 unique ``crashes'', respectively, during this time (Polyglot's efficiency is reduced with semantic validation enabled, resulting in fewer detected crashes). These ``crashes'' are not the result of sanitizer aborts, rather they are all ``Fatal errors'' from the PHP interpreter. 
However, such fatal errors can be intentional rather than true bugs (\texteg if the test case has an expected abort result, it is not an error). We found most ``Fatal errors'' from AFL++ and Polyglot belong to expected aborts. For example, in Figure~\ref{fig:overview}, test B is expected to produce the following failure: ``\code{Fatal error: Uncaught Error: Trying to clone an uncloneable object of class Generator in …}''.
Through manual verification, we confirmed that all ``Fatal errors'' reported by AFL++ and Polyglot are unrelated to memory errors. In contrast, \name identified 16 unique crash sites flagged by sanitizers, primarily related to memory safety issues. \name can also detect all ``Fatal errors'' from the PHP interpreter, however, this would result in a high number of false positives which is undesirable.
% \ry{YC: can we have at te end updated number for AFL++ \& Polyglot removing the fatal error cases?}

% AFL++ and Polyglot detect many more crashes than we count, as they consider the ``Fatal Error'' aborted from the PHP interpreter also as crashes. Nevertheless, those fatal errors, are mostly invalid bugs due to existing error handlers or syntax aborts. The crash sites \name found can cause segmentation fault of the PHP interpreter rather than error aborts. We revisit all crashes reported from these approaches, and execute them in the PHP interpreter without sanitizers, and count the segmentation fault oracle. Our evaluation results show that \name outperforms AFL++ and Polyglot with over 10 unique segmentation faults with different crash sites, while AFL++ and Polyglot detect none but only ``Fatal Errors''. Listing~\ref{lst:5.3} presents one of the segmentation faults found by \name that are not detectable by AFL++ and Polyglot via mutating existing seed programs. \name fuses two tests and fuzzes the internal interfaces (\textie the \code{output\_add\_rewrite\_var()} function) on fused code semantic (\textie the \code{\$x} variable) by randomly calling functions from \code{get\_defined\_functions()} and runtime logging.

\begin{figure*}[t]
\setlength{\abovecaptionskip}{5pt}
\setlength{\belowcaptionskip}{0pt}
\setlength{\intextsep}{0pt}
    \centering
    \includegraphics[scale=0.55]{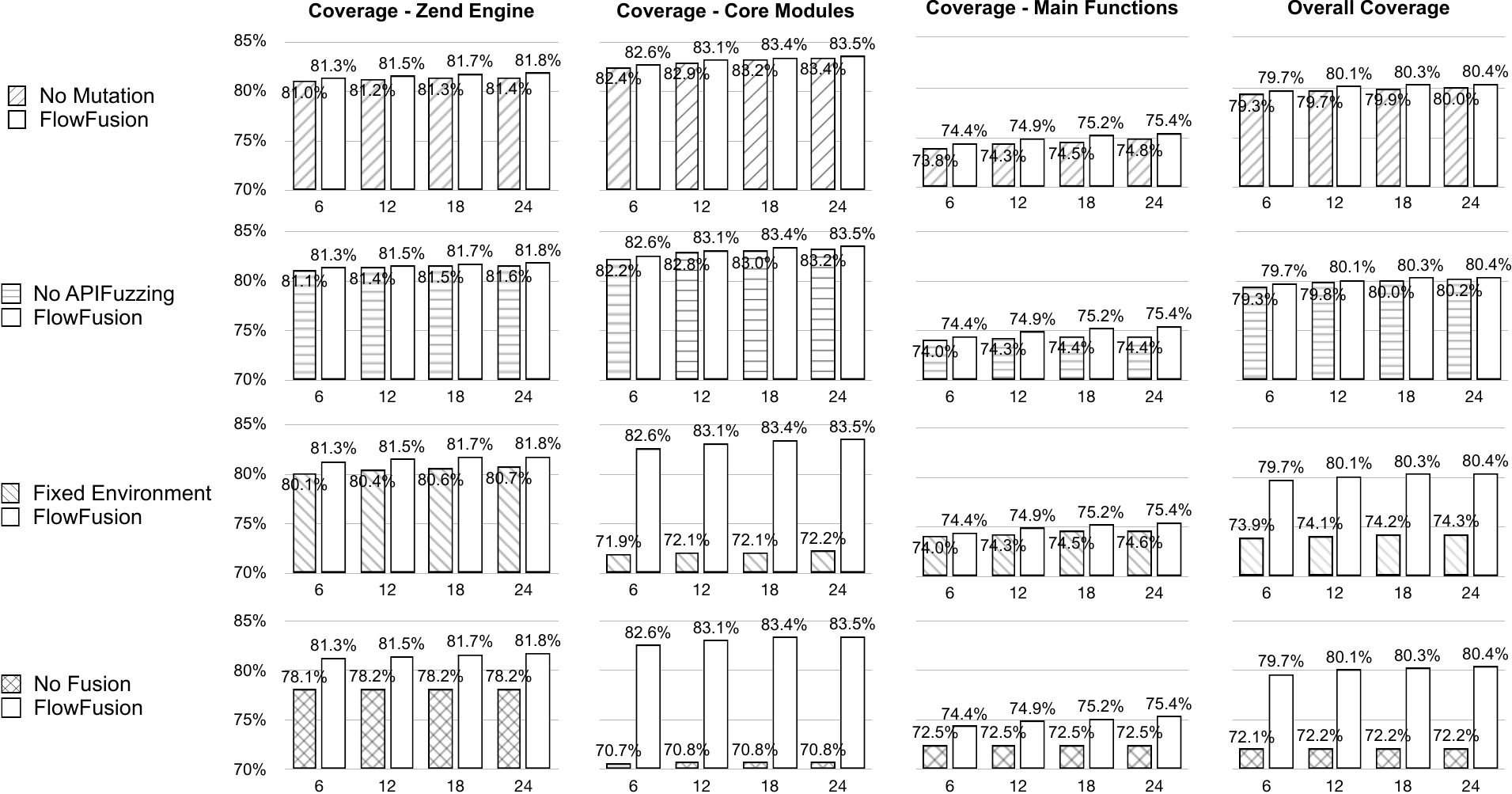}
    \caption{Ablation Study of \name}
    \label{fig:contribution}
\end{figure*}

\vspace{-2mm}
\subsection{Ablation Study}

\label{sec:ablation}

We examined the impact of key strategies used in our approach on overall effectiveness via an ablation assessment. Specifically, we evaluated three strategies introduced in Section~\ref{sec:strategies}: (i) test mutation, (ii) interface fuzzing, (iii) environment crossover, and the key method (iv) dataflow fusion. For this assessment, we configured \name by individually disabling each of these strategies, using a fully enabled version of \name as the reference. To ensure consistency, we compiled five identical copies of the PHP interpreter using the same compiler and options, and evaluated these metrics over 24 hours of fuzzing at the same time, with each configuration limited to 16 parallel threads.

We use code coverage to assess their contribution to effectiveness. We provide detailed coverage data for three main components of the PHP interpreter introduced in Section~\ref{sec:background}: (i) Zend engine, (ii) core modules, and (iii) main functions, along with (iv) the overall coverage. Figure~\ref{fig:contribution} presents the evaluation results with coverage between 70\% to 85\%.
Focusing first on the overall coverage column, we rank the contributions to coverage as follows: dataflow fusion contributes the most with 8\% decrease, followed by environment crossover with 6\% decrease, and finally, test mutation and interface fuzzing with small 0.2\% to 0.4\% decreases. Notably, two significant decreases in code coverage occur with the fixed environment and the absence of dataflow fusion. These strategies are directly related to test fusion—one involving the fusion of program sections and the other merging the rest sections.

We then analyze the three components of the PHP interpreter in relation to overall code coverage. Test mutations result in minor increases across all components, with the Zend engine and main functions showing slightly greater gains than the core modules. Interface fuzzing noticeably boosts coverage in the main functions. Environment crossover leads to a significantly larger increase in the code modules compared to other components. Notably, dataflow fusion drives substantial growth across all components, underscoring its key role in enhancing coverage.

% normal: 16 crash site

% \begin{table}[htbp]
% \centering
% \caption{Unknown Bugs \name Found in PHP interpreter}
% \begin{adjustbox}{width=0.48\textwidth}
% \begin{tabular}{cccccccccccccccc}
% \toprule
% Option  & \multicolumn{3}{c}{No Mutation}                          & \multicolumn{3}{c}{No APIFuzz}                           & \multicolumn{3}{c}{Fixed Env}                            & \multicolumn{3}{c}{No Fusion}                            & \multicolumn{3}{c}{FlowFusion}                           \\ \hline
% Time    & \multicolumn{1}{c}{8h}  & \multicolumn{1}{c}{16h} & 24h & \multicolumn{1}{c}{8h}  & \multicolumn{1}{c}{16h} & 24h & \multicolumn{1}{c}{8h}  & \multicolumn{1}{c}{16h} & 24h & \multicolumn{1}{c}{8h}  & \multicolumn{1}{c}{16h} & 24h & \multicolumn{1}{c}{8h}  & \multicolumn{1}{c}{16h} & 24h \\ \hline
% Crash\# & \multicolumn{1}{c}{123} & \multicolumn{1}{c}{135} & 147 & \multicolumn{1}{c}{123} & \multicolumn{1}{c}{135} & 147 & \multicolumn{1}{c}{123} & \multicolumn{1}{c}{135} & 147 & \multicolumn{1}{c}{123} & \multicolumn{1}{c}{135} & 147 & \multicolumn{1}{c}{123} & \multicolumn{1}{c}{135} & 147 \\
% \bottomrule
% \end{tabular}
% \end{adjustbox}
% \end{table}

\section{Discussion}

\label{sec:discussion}

\vspace{-1mm}

We present several limitations of \name and discuss possible solutions or future works.

% \ry{this is NOT a limitation, I suggested delete, the test suite is ortogonal to grammar issues}

\vspace{+2mm}
\textbf{Efficiency}. The efficiency of \name has not been optimized for exploring larger search spaces. 
% We prioritize efficiency over accuracy by abstracting coarse-grained dataflow from the finer-grained taint tracking used in vulnerability detection~\cite{ji2022flowmatrix}.
We intentionally allow semantic violations when fusing two test cases (\texteg type mismatches or undefined variables). We believe it offers an additional opportunity to uncover more memory errors, some of which developers have remarked as ``\textit{a bit `nonsensical' but should not cause a crash}'' type bug. Introducing semantic validation could be a potential optimization for \name. This would involve more extensive static and dynamic analysis to guide more accurate test fusion, effectively narrowing the search space, and thereby improving performance but potentially missing some edge cases.
% within a reasonable scope, 

\vspace{+2mm}
\textbf{Feedback}. We initially considered incorporating coverage feedback into our approach but ultimately decided against it due to the substantial overhead it introduced. The basic concept was to evaluate each fused test case and add it to the test corpus if it covered more lines of the PHP interpreter. This would involve monitoring code coverage after every execution of fused test cases, which resulted in excessive overhead. We leave this optimization for future work.

\vspace{+2mm}
\textbf{Bug oracle}. We use sanitizers as our primary bug oracle; however, additional bug oracles can be incorporated into our approach to detect further bugs like logic bugs.
%For instance, the official \tests include expected results or failures, serving as an oracle for identifying logic bugs in the PHP interpreter. We plan to implement additional test oracles in future work.
% \ry{removed inconsistency - failure is expected and not a bug so cannot be oracle as argued earlier, this setennce may cause confusion}

\vspace{+2mm}
\textbf{Scalability to other programming languages}. The high-level insight of \name (dataflow interleaving) and other strategies (test mutation, interface fuzzing, and environment crossover) are extensible to other programming languages like C/C++ or JavaScript. We have initial work extending \name to JavaScript, and while it is early, it has already found JavaScript bugs.

\vspace{+2mm}
\textbf{Dependency on the test cases}. 
Relying on test cases benefits \name by automatically reflecting any updates made by developers, but it also constrains \name's scalability---requiring a well-maintained test suite of high quality and ample coverage.

% \name is constrained by the quality and quantity of the existing test suite. Some existing approaches utilize grammar-guided techniques to generate new valid programs, thereby expanding the test suite for fuzzing language interpreters. Our work can be combined with them to incorporate fused tests into the generated test suites, adding additional code semantics and contexts.

\section{Related Work}

\label{sec:related_work}

% \vspace{-1mm}

In this section, we examine related work on PHP application security, historical information reuse, and compiler/interpreter testing and fuzzing to underscore the importance of analyzing the PHP interpreter and to establish the foundation of our approach, \name.

% In this section, we first examine the related work on PHP application security and historical bug or test information reuse to highlight the importance of analyzing the PHP interpreter and pinpoint our 

% to pinpoint the importance of analyzing the PHP interpreter.
% Subsequently, we review existing historical test case reuse strategies to illustrate the generality of our approach.

\vspace{+2mm}
\textbf{PHP application security}.
% Undoubtedly, PHP is one of the most popular languages in web deployments, thus attracting tons of research attention.
% We summarize related works that mitigate web attacks, automatically find vulnerabilities, propose additional defenses, or have intensive evaluation in PHP applications.
% Due to the large number of works on PHP application security, we list parts of the approaches proposed in very recent years.
As one of the most popular languages for web deployments, the security problems in PHP-related applications have recently attracted increasing attention.
To protect these applications, existing solutions focus on attack surface mitigation~\cite{azad2023animatedead,jahanshahi2023minimalist,azad2019less}, bug detection~\cite{al2023whip,neef2024all,luo2022tchecker,rabheru2021deeptective,li2021lchecker}, and defense mechanism enhancement~\cite{bulekov2021saphire,li2011webshield,reis2007browsershield}.
For example, Minimalist~\cite{jahanshahi2023minimalist} and AnimateDead~\cite{azad2023animatedead} use debloating strategies to reduce the code size of applications, such as minimizing critical API calls, thereby increasing the complexity and workload for attackers.
WHIP~\cite{al2023whip} enables static application security testing (SAST) tools to collaborate by sharing information, which helps to trigger more security alerts.
TChecker~\cite{luo2022tchecker} introduces a context-sensitive inter-procedural static taint analysis tool to detect taint-style vulnerabilities in PHP applications.
Additionally, Saphire~\cite{bulekov2021saphire} applies the principle of least privilege (PoLP) to PHP applications and proposes a novel, generic approach for automatically deriving system-call policies for individual interpreted programs.
BrowserShield~\cite{reis2007browsershield} utilizes a lightweight middlebox to prevent the exploitation of browser vulnerabilities. \name is orthogonal to these works, aiming to detect memory errors that provide an additional low-level attack surface alongside common application-level bugs.

\vspace{+2mm}
\textbf{Reusing historic bugs or tests}.
Historical bug reports and test cases offer valuable insights into a system's internal state and potentially vulnerable points.
These insights are widely used to construct new, meaningful inputs or seeds that help uncover additional vulnerabilities.
For example, Oliinyk et al.~\cite{oliinyk2024fuzzing} and Zhao et al.~\cite{zhao2022history} analyze previous crash or bug reports to facilitate new seed generation, identifying new bugs in BusyBox and the Java Virtual Machine (JVM), respectively.
Alternatively, some approaches~\cite{zhong2022enriching,su2025understanding,le2014compiler,le2015finding,sun2016finding} leverage existing test cases as guidance to explore program states and discover new vulnerabilities.
For instance, SQuaLity~\cite{su2025understanding} executes test suites across different Database Management Systems (DBMSs) to uncover new, previously unknown bugs.
In the context of C/C++ compiler testing, the Equivalence Modulo Inputs (EMI) approach~\cite{le2014compiler, le2015finding, sun2016finding} mutates existing test programs to create semantically equivalent variants, aiding in bug discovery.
Additionally, YinYang~\cite{winterer2020validating} goes a step further by generating new test inputs through semantic fusion, combining two existing formulas into a new one to detect soundness bugs in Satisfiability Modulo Theory (SMT) solvers. Guided by this, \name explores the potential of fusing high-quality test cases to discover PHP memory errors, enhancing the security of the PHP interpreter.

% \yc{Add David Lie LLVM paper FLUX: Finding Bugs with LLVM IR Based Unit Test Crossovers} 

% FLUX~\cite{liu2023flux}
% Several notable works have explored reusing historic bugs or tests to uncover additional bugs in widely used software and systems. 
% One such approach~\cite{zhao2022history} involves generating new test programs by extracting historical bug programs into seed programs for finding bugs in the Java Virtual Machine (JVM). 
% YinYang~\cite{winterer2020validating}, is for discovering soundness bugs in Satisfiability Modulo Theory (SMT) solvers. 
% YinYang utilizes Semantic Fusion, which merges two existing formulas into a new one, ensuring satisfiability is preserved by construction. 
% LeRE~\cite{zhong2022enriching} extracts and reuses test programs from bug reports to detect compiler bugs with extracted programs.
% SQuaLity~\cite{su2025understanding} demonstrates the benefits of reuse by executing test suites across different Database Management Systems (DBMSs). 
% Mutation testing, which involves altering existing tests or programs to generate new ones, represents another related direction. 
% For instance, in C/C++ compiler testing, the Equivalence Modulo Inputs (EMI) approach~\cite{le2014compiler, le2015finding, sun2016finding} mutates existing test programs to produce semantically equivalent variants, aiding in bug discovery.

\vspace{+2mm}
\textbf{Compiler/Interpreter testing and fuzzing}.
Existing solutions have concentrated on fuzzing or testing popular compilers and interpreters, such as those for C/C++\cite{yang2011finding,even2023grayc,li2024boosting}, Rust\cite{sharma2023rustsmith,tuong2023symrustc}, and JavaScript (JS)~\cite{gross2023fuzzilli,wang2023fuzzjit,bernhard2022jit}, to mitigate potential cascading security issues.
For instance, GrayC~\cite{even2023grayc} design a greybox, coverage-directed, mutation-based approach to fuzz C compilers and code analyzers using a new set of mutations to target common C constructs. Creal~\cite{li2024boosting} boosts C/C++ compiler testing by fusing real-world code with seed programs. 
RustSmith~\cite{sharma2023rustsmith} executes differential testing between Rust compilers or across optimization levels to identify potential bugs.
Comfort~\cite{ye2021automated} leverages the deep learning-based language model to automatically generate JS test code to detect bugs in JS engines.
Fuzzilli~\cite{gross2023fuzzilli} presents the design and implementation of an intermediate representation (IR) aimed at uncovering vulnerabilities in JIT compilers.
FuzzJIT~\cite{wang2023fuzzjit} focuses on identifying JIT compiler bugs by triggering the JIT compilation process and capturing execution inconsistencies. 
CodeAlchemist~\cite{han2019codealchemist} leverages syntax-aware assembly by merging code bricks from the seed to generate new test cases for fuzzing JavaScript engines.
% Their semantics-aware assembly is also a form of dataflow fusion, the difference is that they are specific to JavaScript whereas \name is tailored to PHP---our ablation study shows that not only is dataflow important but also interface fuzzing and environment crossover.
% \ry{YC: added more on CodeAlchemist}
LangFuzz~\cite{holler2012fuzzing}, NAUTILUS~\cite{aschermann2019nautilus}, Gramatron~\cite{srivastava2021gramatron}, and PolyGlot~\cite{chen2021one} are existing fuzzing approaches that have found memory errors in the PHP interpreter. Focusing solely on grammar may overlook the code semantics of programs, thereby limiting their bug-discovery capabilities. To address this, we developed \name to detect memory errors in the PHP interpreter by fusing dataflow from high-quality test cases, generating new and more complex code semantics.

% \vspace{-1mm}

% There is a vast body of work on compiler testing and fuzzing. Here, we highlight some of the most recent papers specifically focused on C/C++ and JavaScript compilers.

% GrayC~\cite{even2023grayc} introduces a graybox, coverage-directed, mutation-based approach for fuzzing C compilers and code analyzers.
% ComFuzz~\cite{ye2023generative} enhances the efficiency of compiler fuzzing by targeting specific components and features that are more likely to uncover compiler bugs.
% Comfort~\cite{ye2021automated} leverages deep learning to detect bugs in JavaScript engines. 
% FuzzJIT~\cite{wang2023fuzzjit} focuses on identifying JIT compiler bugs by triggering the JIT compilation process and capturing execution inconsistencies. Jit-Picking~\cite{bernhard2022jit} uses differential fuzzing to reveal JIT bugs. \yc{TODO: add php-related grammar fuzzers: LangFuzz, Nautilus, Gramatron, Polyglot} NAUTILUS~\cite{aschermann2019nautilus}, Gramatron~\cite{srivastava2021gramatron} find some memory errors in the PHP interpreter via grammar guidance. \yc{TODO: one sentence to summarize our difference}

\section{Conclusion}

\label{sec:conclusion}

% \vspace{-1mm}

In this paper, we introduced \name, the first automated fuzzing framework specifically designed to detect memory errors in the PHP interpreter through dataflow fusion and other innovative techniques. Our approach merges test cases by linking their dataflows, creating fused tests capable of uncovering previously undetected bugs. Comprehensive experiments demonstrated \name's effectiveness, revealing \totalbug bugs, with \fixbug successfully fixed and \confbug confirmed, outperforming existing methods. We believe that continued fuzzing with \name will discover even more PHP bugs.

\name demonstrates its potential as a practical tool for improving the security and robustness of the PHP interpreter. Moreover, the principles behind \name’s design—especially the dataflow fusion technique detailed in Algorithm~\ref{alg:fusion}—are not limited to PHP.
% PHP-specific features. This suggests 
We believe there is broader applicability to other programming languages, where \name could be effectively used to merge test cases based on dataflow relationships, making \name a valuable contribution to the field of language interpreter fuzzing.
% We believe this \name has practical extensibility and generalizability, which allow it to fuzz many languages including interpreted and dynamic ones where the language runtime needs to be factored in.
% underscores the versatility and generalizability of our method, making \name a valuable contribution to the field of language interpreter fuzzing.
% \ry{updated}

% \clearpage

\section*{Ethics Statement}

In conducting this research, we have carefully considered the ethical implications at every stage, from design through publication. We avoided live experimentation on systems without proper authorization and ensured that any vulnerabilities identified during the research were responsibly disclosed to relevant parties. By following these ethical practices, we aim to minimize harm and ensure that our research contributes positively to the field. These considerations not only align with ethical standards but also promote responsible innovation in the computer security and privacy domains. 

\section*{Open Science Statement}

We are committed to openly sharing all research artifacts associated with this work. Our approach, \name, is available at \url{https://github.com/php/flowfusion} under an open-source license and archived paper artifact at \url{https://zenodo.org/records/14642350}. Our commitment to open science aligns with the broader initiative to foster transparency and collaboration within the research community.

\section*{Acknowledgements}

We appreciate the PHP developers’ responsiveness in addressing our bug reports, providing valuable feedback, and cooperating with us to open source \name. This research is supported by the National Research Foundation, Singapore, and Cyber Security Agency of Singapore under its National Cybersecurity R\&D Programme (Fuzz Testing <NRF-NCR25-Fuzz-0001>) and by MOE grant A-8001544-00-00. Any opinions, findings and conclusions or recommendations expressed in this material are those of the author(s) and do not reflect the views of National Research Foundation, Singapore and Cyber Security Agency of Singapore.

{\footnotesize\balance \bibliographystyle{acm}
\bibliography{sample}}

% \clearpage

% \section{Appendix}

\end{document}